\documentclass[usenatbib]{mn2e}

\usepackage{amsmath}
\usepackage{natbib}
\usepackage{epsfig}
\usepackage{txfonts}

\newcommand{\id}{{\,\rm d}}

\newcommand{\beq}{\begin{equation}}   %

\newcommand{\eeq}{\end{equation}}   %

\newcommand{\beqa}{\begin{eqnarray}}   %

\newcommand{\eeqa}{\end{eqnarray}}   %

\newcommand{\beal}{\begin{align}}

\newcommand{\enal}{\end{align}}

\newcommand{\bspl}{\begin{split}}

\newcommand{\espl}{\end{split}}

\newcommand{\bsub}{\begin{subequations}}

\newcommand{\esub}{\end{subequations}}

\newcommand{\bmulti}{\begin{multline}}   %

\newcommand{\beqm}{\begin{mathletters}}   %

\newcommand{\eeqm}{\end{mathletters}}   %

\newcommand{\Abl}[2]{\frac{{\rm d} #1}{{\rm d} #2}}

\newcommand{\pot}[2]{#1 \times 10^{#2}}

\newcommand{\ion}[2]{{\text{{\sc #1}\,{\sc #2}}}}
\newcommand{\HeIlevel}[4]{{#1^{#2} {\rm #3}_{#4}}}   % \HeIlevel{n}{2S+1}{L}{J}

\usepackage{color}

\title[Could the Cosmological Recombination Spectrum Help Us Understand Annihilating Dark Matter?]
{Could the Cosmological Recombination Spectrum Help Us Understand Annihilating Dark Matter?}

\author[Chluba]{J. Chluba$^{1,2}$\thanks{E-mail: jchluba@cita.utoronto.ca} 
  \\
$^{1}$ Canadian Institute for Theoretical Astrophysics, 60 St. George Street,
Toronto, ON M5S 3H8, Canada\\
$^{2}$ Max-Planck Institut f\"ur Astrophysik, Karl-Schwarzschild-Str. 1,
D-85740 Garching, Germany\\
}

\begin{document}

\date{Received **insert**; Accepted **insert**}

\maketitle

\begin{abstract}
In this paper we explore the potential effects of DM annihilations on the {\it cosmological recombination spectrum}.
With this {\it example} we want to demonstrate that the cosmological recombination spectrum in principle is {\it sensitive} to details related to possible extra energy release during recombination. 
We restrict ourselves to DM models which produce a {\it negligible primordial distortion} of the CMB energy spectrum (usually characterized as $\mu$ and $y$-type distortions). 
However, since during the epoch of cosmological recombination ($z\sim 1000$) a large fraction of the deposited energy can {\it directly} go into ionizations and excitations of neutral atoms, both the cosmological recombination spectrum and ionization history can still be affected significantly.
%%
%Motivated by the anomalies in the electron and positron cosmic-ray spectra seen by {\sc Pamela}, {\sc Fermi}, {\sc Hess} and {\sc Atic}, the possible effects %of DM annihilations on the ionization history of the Universe have recently been addressed by several groups using modified versions of {\sc Recfast}. 
%%
%Here we now also compute the modifications in the CMB spectral distortions produced during the cosmological recombination epoch with our full multi-%level \ion{H}{i} and \ion{He}{i} recombination code.
%%
We compute the modifications to the cosmological recombination spectrum using our multi-level \ion{H}{i} and \ion{He}{i} recombination code, showing that  {\it additional} photons are created due to uncompensated loops of transitions which are induced by DM annihilations.
As we illustrate here, the results depend on the detailed branching of the deposited energy into heating, ionizations and excitations. This dependence in principle should allow us to shed light on the nature of the underlying annihilating DM model (or more generally speaking, the mechanism leading to energy injection) when measuring the cosmological recombination spectrum. 
However, for current upper limits on the potential DM annihilation rate during recombination the cosmological recombination spectrum is only affected at the level of a few percent. 
Nevertheless, we argue here that the cosmological recombination spectrum would provide another independent and very direct way of checking for the presence of sources of extra ionizing or exciting photons at high redshifts. This would open an new window to possible (non-standard) processes occurring {\it before}, {\it during} and {\it between} the {\it three} epochs of recombination.
% at $\sim 230\,000$, $\sim130\,000$ and $\sim 18\,000$ years after the big bang.
%
\end{abstract}
%------------------------------------------------------------

\begin{keywords}
Cosmic Microwave Background: cosmological recombination, temperature
  anisotropies, cosmological recombination spectrum, spectral distortions
\end{keywords}

\section{Introduction}
\label{sec:Intro}
%---------------
The anomalies in the cosmic-ray spectra of electrons and positrons at $1\,{\rm GeV}-1000\,{\rm GeV}$ seen using {\sc Pamela} \citep{Adriani2009Nature}, {\sc Fermi} \citep{Abdo2009}, {H.E.S.S.} \citep{Hesscoll2009}, and {\sc Atic} \citep{Chang2008} could be interpreted as signatures from {\it annihilating dark matter} \citep[e.g. see][and references therein]{Hooper2004, Cholis2008, Hamed2009}.
This very intriguing possibility has recently motivated several independent groups to reconsider the effects of such dark matter (DM) models on the ionization history of our Universe and the cosmic microwave background (CMB) temperature and polarization anisotropies \citep{Galli2009, Slatyer2009, Cirelli2009, Huetsi2009, Kanzaki2009} using modified versions of {\sc Recfast} \citep{SeagerRecfast1999, Seager2000, Wong2008}. 

It is clear that the ionization history is sensitive to the number of extra {\it ionizations} and {\it excitations} of neutral hydrogen or helium atoms at $z\sim 1000$ \citep[e.g.][]{Peebles2000}. 
These extra ionizations or excitations can be mediated by {\it particles} (e.g. electrons or positrons) or {\it photons}, which, for example, could be produced as a consequence of DM annihilations \citep[e.g.][]{Padmanabhan2005} or {\it decaying particles} \citep[e.g.][]{Chen2004}. 
Therefore, both DM annihilations or decays of long-lived unstable particles in principle are able to {\it delay} recombination, introducing changes to the Thomson visibility function \citep{Sunyaev1970} at $z\sim 1100$, which then affect the cosmic microwave background (CMB) temperature and polarization anisotropies, changing the position of the acoustic peaks and their (relative) heights.
This allows one to place interesting constraints on possible DM annihilations during recombination using current and future CMB data \citep[e.g.][]{Zhang2006, Galli2009, Slatyer2009}. 
Similarly, energy release by decaying particles \citep{Zhang2007}, or more general sources of additional ionizations or excitations of neutral atoms during recombination \citep{Bean2003, Bean2007, Galli2008} can be constraint.
Since the presence of any such sources could compromise our ability to measure the spectral index of the primordial power spectrum and its running, it is very important to consider these possibilities carefully, along with other {\it physical} corrections to the modeling of cosmological recombination \citep[see][for detailed overview on recently considered processes]{Fendt2009, Sunyaev2009}.

However, the CMB temperature and polarization anisotropies are not the only (direct) signals that can tell us about (non-standard) processes occurring during cosmological recombination.
It is well known that the recombination of hydrogen and helium in the Universe leads to the emission of several photons per baryon, modifying the CMB energy spectrum \citep{Zeldovich68, Peebles68, Dubrovich1975, Dubrovich1997}.  
Recently, detailed computations of the cosmological recombination spectrum were carried out \citep[e.g.][]{Jose2006, Chluba2006, Jose2008}, showing that the recombinations of hydrogen and helium lead to relatively {\it narrow spectral features} in the CMB energy spectrum. These features were created at redshift $z\sim 1300-1400$, $\sim 2100-2400$ and $\sim 6000$, corresponding to the times of \ion{H}{i}, \ion{He}{i} and \ion{He}{ii} recombination, and, due to redshifting, today should still be visible at mm, cm and dm wavelength. Observing these signatures from cosmological recombination may offer an independent way to determine some of the key cosmological parameters, such as the {\it primordial helium abundance}, the {\it number density of baryons} and the CMB {\it monopole temperature} at recombination \citep[e.g.][]{Chluba2008T0}. Furthermore, it will allow us to {\it directly} check our understanding of the recombination process and possible non-standard aspects \citep[e.g. see][for an overview]{Sunyaev2009}, for example, in connection with early energy release \citep{Chluba2008c}.

In this work we will {\it demonstrate} that the cosmological recombination spectrum also is sensitive to the branching of energy released due to DM annihilations into ionizations and excitations. As we show here, it is not only important how much energy is deposited in total, but also when.
Depending on the underlying model for the annihilating DM these efficiencies will differ, so that observing the cosmological recombination spectrum may offer another very direct way for constraining such models.

In earlier considerations of the possible effects in connection with energy release during recombination, this branching was either parametrized with single numbers \citep{Peebles2000, Bean2003, Bean2007}, or simple approximations were used \citep{Chen2004, Padmanabhan2005, Mapelli2006}.
A detailed account for all possible aspects of the problem related to a computation of these efficiencies is beyond the scope of this paper. Based on earlier investigations, we will therefore restrict ourselves to some simple examples (see Sect.~\ref{sec:modelling}), which are mainly intended to show the principle dependencies of the recombination spectrum on DM annihilations. 
However, one can carry out similar computations in connection with decaying particles or other, more speculative sources of extra ionizations during  recombination. Since in those cases the cosmological recombination should also exhibit signatures of these non-standard processes, we hope that this work will provide further motivation towards refined studies in connection with the cosmological recombination spectrum and potential future experiments measuring it in detail.

{\it But how do additional ionizations or excitations during recombination actually affect the cosmological recombination spectrum?}
Every extra ionization of hydrogen liberates an electron and proton. At $z\sim1000$ this process is reversed by a recombination of the proton with another free electron after a rather short time. Since the \ion{H}{i} Lyman continuum is completely blocked \citep{Chluba2007b}, the electron is captured into some excited state ($n\geq 2$), emitting at least two photons in the subsequent cascade towards the ground-state.
This increases the total emission of photons by hydrogen during recombination, since in every additional {\it loop} of transitions several quanta can be produced.
Similarly, extra excitations allow additional electrons to reach high levels, so that in total more recombination photons will be released, by both hydrogen and at $z\sim 2200$ also by helium.
The physics of this problem is very similar to the effect of helium photons on hydrogen \citep{Chluba2009c}, leading to additional feedback-induced emission during recombination.
Here the source of extra ionizations and excitations is related to photons emitted by helium at $z\sim 2200$ and $\sim 6000$ in the normal recombination process.
Another example is connected with the changes introduced to the cosmological recombination spectrum as a consequence of early energy release, which produces an primordial $y$-type distortion \citep{Zeldovich1969} of the CMB in the pre-recombinational epoch, leading to uncompensated loops of atomic transitions that attempt to restore the CMB blackbody spectrum \citep{Chluba2008c}. 
In this case, the extra ionizations and excitations are caused by the excess of photons in the Wien tail of the distorted CMB.

For the cases considered here the latter process is not important, since the required energy that goes into additional ionizations or excitations during recombination can be {\it tiny} in comparison with the energy density of the CMB.
In such cases the CMB spectrum is not affected significantly by the additional energy release throughout the entire pre-recombinational epoch (see Appendix~\ref{sec:est_y} for some estimates), while the dynamics of cosmological recombination can still be strongly modified.
This is because the number of hydrogen and helium nuclei is a factor of $\sim \pot{2}{10}$ smaller than the number of CMB photons. This makes it rather easy to perturb the recombination process, while at the same time the CMB energy spectrum itself remains practically unaltered.

%The paper is structured as follows: in Sect.~\ref{sec:modelling} we provide some additional details on the modelling of the processes in connection with %annihilating DM and the recombination process. There we also explain some of the 

\section{Modeling of the different processes}
\label{sec:modelling}
%---------------
In this section we give a brief summary of the required equations for our multi-level recombination code to take the effect of DM annihilations into account. We closely follow the approach outlined by \citet{Chen2004} in connection with energy injection from decaying particles, including recent modifications and updates related to DM annihilations \citep{Padmanabhan2005, Galli2009, Slatyer2009, Huetsi2009}.
However, some of the details here are slightly different, and we also introduce the basis for further improvements of our multi-level recombination code in connection with this problem.

\subsection{Overall energy injection rate}
\label{sec:dE_dt}
%-----------
Envisioning some self-annihilating DM particle $\chi$ and its antiparticle $\bar{\chi}$, the total rate of energy release per unit volume is given by
%-----------
\bsub
\label{eq:dE_dt}
\beal
\label{eq:dE_dt_a}
\left.\Abl{E}{t}\right|_{\rm \chi\bar{\chi}}
&=2\,M_\chi c^2 \left<\sigma v\right> \,N_\chi\,N_{\bar{\chi}}
\\[1mm]
\label{eq:dE_dt_b}
&\approx\pot{2.9}{-31}\,[1+z]^6 \,{\rm eV\,s^{-1}\,cm^{-3}}
\nonumber\\
&\qquad\quad\times
\left[\frac{M_\chi c^2}{100\,{\rm GeV}}\right]^{-1}
\left[\frac{\Omega_{\chi} h^2}{0.13}\right]^2 
\left[\frac{\left<\sigma v\right>}{\pot{3}{-26}\,{\rm cm^3/s}}\right] 
%\\[1mm]
%\label{eq:dE_dt_c}
%&\approx\pot{1.5}{-24}\,N_{\rm H}\,[1+z]^3 \,{\rm eV\,s^{-1}}
%\nonumber\\
%&\qquad\quad\times
%%\left[\frac{\Omega_{\rm b} h^2}{0.022}\right]^{-1} 
%%\left[\frac{1-Y_{\rm p}}{0.76}\right]^{-1} 
%\left[\frac{M_\chi c^2}{100\,{\rm GeV}}\right]^{-1}
%\left[\frac{\Omega_{\chi} h^2}{0.13}\right]^2 
%\left[\frac{\left<\sigma v\right>}{\pot{3}{-26}\,{\rm cm^3/s}}\right] 
\end{align}
\esub
%-----------
Here $M_\chi \equiv M_{\bar{\chi}}$ is the mass of the DM particle and its antiparticle; $\left<\sigma v\right>$ is the thermally
averaged product of the cross-section and relative velocity of the annihilating dark matter particles; and $N_\chi\equiv N_{\bar{\chi}}=N_{0,\chi} [1+z]^3$ is the number density of DM particles and their antiparticles, with the present day value $N_{0,\chi}\approx \pot{1.4}{-8}\,{\rm cm^{-3}}\,
\left[\frac{\Omega_{\chi} h^2}{0.13}\right] \left[\frac{2\,M_\chi c^2}{100\,{\rm GeV}}\right]^{-1}$. 
%%
%In Eq.~\eqref{eq:dE_dt_c} we also used $N_{\rm H}\approx \pot{1.9}{-7}\,{\rm cm^{-3}}[1+z]^3$ for the number density of hydrogen nuclei in the Universe.

Depending on the DM model, $\left<\sigma v\right>$ in general is a function of redshift. In particular, it could also include the effect of Sommerfeld enhancement during the epoch of cosmological recombination \citep[e.g.][]{Galli2009, Slatyer2009}, which can be important since at that time the relative velocities of the DM particles become small.
One can incorporate these possibilities by replacing $\left<\sigma v\right>=S(z)\,\left<\sigma v\right>_{0}$, where $\left<\sigma v\right>_{0}={\rm const}$, however below we will restrict ourselves to cases with $\left<\sigma v\right>=S(z)\,\left<\sigma v\right>_{0}={\rm const}$.

Another aspect of the problem is connected with the {\it clustering} of DM \citep{Huetsi2009}. This effect becomes only important at low redshift ($z\lesssim 100$) and leads to an enhancement $\left<N_\chi\,N_{\bar{\chi}}\right>=B(z) \left<N_\chi\right>^2$ of the average squared DM number density, with clustering boost factor $B(z)>1$.
This effect can become very pronounced, depending on the assumed halo concentration model and the lower mass cutoff for the halo mass function. However, here we are mainly interested at the CMB spectral distortions generated at $z\gtrsim 200$, where the clustering of matter is negligible.
We therefore neglect this aspect of the problem here, using $B(z)=1$ throughout, so that Eq.~\eqref{eq:dE_dt} remains unaltered.

However, it is very important that depending on the involved annihilation channels (e.g. photons, leptons, hadrons, neutrinos) only a fraction, $f_{\rm d}$, of the released energy will be {\it deposited} into the intergalactic medium (IGM), going into {\it heating}, and {\it ionizations} or {\it excitations} of atoms (i.e. hydrogen and helium). 
For example, energy released in form of neutrinos (at redshifts of interest to us here) will be carried away, so that usually one expects $f_{\rm d}<1$.
Furthermore, because the {\it transparency} of the Universe to photons and the energy deposition efficiency of different particles (e.g. electrons and positrons) depend on the redshift of injection and the cosmological model (e.g. densities and expansion rate),
$f_{\rm d}$ is a function of time and cosmology.

To include this aspect of the problem into the computations we therefore write
%-----------
\beal
\label{eq:dEd_dt_}
\left.\Abl{E_{\rm d}}{t}\right|_{\rm \chi\bar{\chi}}
&=f_{\rm d}(z)\,\left.\Abl{E}{t}\right|_{\rm \chi\bar{\chi}}
=f_{\rm d}(z)\,{\epsilon}_0 \,N_{\rm H}\,[1+z]^3 \,\rm eV\,s^{-1},
\end{align}
%-----------
with the dimensionless parameter
%-----------
\beal
\label{eq:deps_dt_}
{\epsilon}_0=\pot{1.5}{-24}\,
\left[\frac{M_\chi c^2}{100\,{\rm GeV}}\right]^{-1}
\left[\frac{\Omega_{\chi} h^2}{0.13}\right]^2 
\left[\frac{\left<\sigma v\right>}{\pot{3}{-26}\,{\rm cm^3/s}}\right]. 
\end{align}
%-----------
Here we also used $N_{\rm H}\approx \pot{1.9}{-7}\,{\rm cm^{-3}}[1+z]^3$ for the number density of hydrogen nuclei in the Universe.

A detailed computation for $f_{\rm d}$ as a function of time and cosmology is beyond the scope of this paper.
However, recently \citet{Slatyer2009} computed the function $f_{\rm d}$ for different models of annihilating DM for the concordance model. 
Their models give typical values for $f_{\rm d}$ ranging from $f_{\rm d}\sim 0.1$ to $f_{\rm d}\sim 1$, where the largest values are reached for annihilation channels $\chi\bar{\chi}\rightarrow e^+ e^-$ at high ($z\gtrsim 10^3$) redshifts. 
They also provided some simple fitting formulae for specific DM models, which we will use below. For the purpose of this paper this should be sufficient.

\subsection{Heating of the medium}
\label{sec:Heating}
%---------------
As mentioned above, only part of the energy that is deposited into the IGM will go into heating of the medium. The rest will lead to ionizations or excitations of hydrogen and helium atoms.
If we denote the fraction of the deposited energy that goes into heating by $g_{\rm h}(z)$, then we can write the additional term in the evolution equation of the temperature of the medium which is related to DM annihilations as
%-----------
\beal
\label{eq:dT_dt}
\left.\Abl{T_{\rm M}}{t}\right|_{\rm h}
%=\frac{f_{\rm h}(z)}{\alpha(z)}\,\left.\Abl{E}{t}\right|_{\rm \chi\bar{\chi}}
=\frac{2}{3k}\,\frac{g_{\rm h}(z)}{N_{\rm H}[1+f_{\rm He}+X_{\rm e}]}\,\left.\Abl{E_{\rm d}}{t}\right|_{\rm \chi\bar{\chi}}.
\end{align}
%-----------
Here $N_{\rm H}$ is the total number of hydrogen nuclei; $f_{\rm He}\sim 8\%$ is the number of helium nuclei relative to the number of hydrogen nuclei; and $X_{\rm e}=N_{\rm e}/N_{\rm H}$ is the usual free electron fraction.

It is clear that at high redshifts, well before the epoch of recombination, practically all the deposited energy goes into heating of the medium, so that $g_{\rm h}(z)\sim 1$.
Due to energy conservation, one also has
$g_{\rm h}(z)= g_{\rm d}(z)-g_{\rm ion}(z)-g_{\rm ex}(z)$,
%%-----------
%\beal
%\label{eq:f_h}
%f_{\rm h}(z)= f_{\rm d}(z)-f_{\rm i}(z)-f_{\rm ex}(z),
%\end{align}
%%-----------
where here $g_{\rm ion}(z)$ is the fraction of the deposited energy that goes into ionizations of atoms, and $g_{\rm ex}(z)$ the fraction that goes into excitations. 
Note that every ionization event also leads to partial heating of the medium, since the liberated electron (and nucleus) will usually have some (large) excess energy which then will be dissipated, e.g. in form of {\it secondary} particles, which again can lead to the heating and ionization of the medium. In detailed computations of the efficiencies $g_{\rm ion}$ and $g_{\rm ex}$ this has to be accounted for. 
Below we will specify which approximations for these functions we use in this work.

\subsection{Ionizations and excitations of atoms}
\label{sec:ion_ex}
%---------------
Knowing that a fraction $g_{\rm ion}(z)$ of the energy deposited by DM annihilations is going into ionizations of both hydrogen an helium, one can write $g_{\rm ion}(z)=g^{\rm H}_{\rm ion}(z)+g^{\rm He}_{\rm ion}(z)$, where $g^{\rm H}_{\rm ion}(z)$ and $g^{\rm He}_{\rm ion}(z)$ are the partial contributions of hydrogen and helium, respectively. Similarly, one has $g_{\rm ex}(z)=g^{\rm H}_{\rm ex}(z)+g^{\rm He}_{\rm ex}(z)$ for the energy that goes into excitations.
Since the chemical mixture of the medium can vary, it is useful to introduce the specific contributions $\tilde{g}^{\rm i}_{\rm a}(z)=g^{\rm i}_{\rm a}(z)\,N_{\rm tot}/N_{\rm i}$, with ${\rm i=H}$ or ${\rm i=He}$; and ${\rm a=ion}$ or ${\rm i=ex}$. For hydrogen one has $N_{\rm tot}/N_{\rm H}=(1+f_{\rm He})$ and for helium $N_{\rm tot}/N_{\rm He}=(1+f_{\rm He})/f_{\rm He}$.

For the net ionization rate from the ground states of neutral hydrogen and helium related to DM annihilations one then finds
%-----------
\bsub
\label{eq:dN_dt_i}
\beal
\label{eq:dN_dt_i_a}
\left.\Abl{N^{\ion{H}{i}}_{\rm 1s}}{t}\right|_{\rm i}
&=-\frac{1}{1+f_{\rm He}}\,\frac{\tilde{g}^{\rm H}_{\rm ion}(z)}{E^{\ion{H}{i}}_{\rm ion}}\,\left.\Abl{E_{\rm d}}{t}\right|_{\rm \chi\bar{\chi}}
\\
\label{eq:dN_dt_i_b}
\left.\Abl{N^{\ion{He}{i}}_{\rm 1s}}{t}\right|_{\rm i}
&=-\frac{f_{\rm He}}{1+f_{\rm He}}\,\frac{\tilde{g}^{\rm He}_{\rm ion}(z)}{E^{\ion{He}{i}}_{\rm ion}}\,\left.\Abl{E_{\rm d}}{t}\right|_{\rm \chi\bar{\chi}},
\end{align}
\esub
%-----------
where $E^{\ion{H}{i}}_{\rm ion}=13.6\,$eV and $E^{\ion{He}{i}}_{\rm ion}=24.6\,$eV are the ionization potentials of hydrogen and helium, respectively.
For Eq.~\eqref{eq:dN_dt_i} it was assumed that \ion{He}{iii} is not important. Furthermore, it is assumed that the energy which is consumed in each ionization is equal to the ionization energy, so that $\frac{g_{\rm ion}}{E_{\rm ion}}\left.\Abl{E}{t}\right|_{\rm \chi\bar{\chi}}$ gives the rate of ionization events per unit volume.  
By definition of $g_{\rm ion}$ this should be possible.

With the same arguments, for excitations of hydrogen and helium from the ground state one can write
%-----------
\bsub
\label{eq:dN_dt_ex}
\beal
\label{eq:dN_dt_ex_a}
\left.\Abl{N^{\ion{H}{i}}_{\rm 1s}}{t}\right|_{\rm ex}
&=-\frac{1}{1+f_{\rm He}}\,\frac{\tilde{g}^{\rm H}_{\rm ex}(z)}{E^{\ion{H}{i}}_{\rm ex}}\,\left.\Abl{E_{\rm d}}{t}\right|_{\rm \chi\bar{\chi}}
\\
\label{eq:dN_dt_ex_b}
\left.\Abl{N^{\ion{He}{i}}_{\rm 1s}}{t}\right|_{\rm ex}
&=-\frac{f_{\rm He}}{1+f_{\rm He}}\,\frac{\tilde{g}^{\rm He}_{\rm ex}(z)}{E^{\ion{He}{i}}_{\rm ex}}\,\left.\Abl{E_{\rm d}}{t}\right|_{\rm \chi\bar{\chi}},
\end{align}
\esub
%-----------
where now $E^{\ion{H}{i}}_{\rm ex}=10.2\,$eV and $E^{\ion{He}{i}}_{\rm ex}\approx 21.0\,$eV are the transition energies to the second shell of hydrogen and helium, respectively. 
Here in particular it is assumed that states with principle quantum numbers $n>2$ are not directly excited. For the purpose of this paper this approximation will do, however, this may lead to an underestimation of the total number of additional secondary low-frequency photons produced in our computations. 
This is because excitations to highly excited levels ($n\gg 2$) will directly allow some electrons to make transitions among excited states, which leads to  emission at low frequencies. On the other hand, for excitations to the second shell the electron will more likely stay within lower levels, and hence produce emission at high frequencies.

\subsubsection{Expressions for $\tilde{g}^{\rm i}_{\rm ion}(z)$ and $\tilde{g}^{\rm i}_{\rm ex}(z)$}
\label{sec:f_ion_ex}
%---------------
Detailed computations of the specific ionization and excitation fractions  $\tilde{g}^{\rm i}_{\rm ion}(z)$ and $\tilde{g}^{\rm i}_{\rm ex}(z)$ require to follow the evolution of primary and secondary, non-thermal electrons and photons produced by the DM annihilation process \citep[see][for a recent study in this connection]{Kanzaki2009}.
Such computations have to include several cooling (e.g. Compton and Coulomb cooling for electrons) and particle creation processes (e.g. pair production by photons), and various aspects of radiative transfer (e.g. photon feedback, escape of photons from the main resonances). 
This is far beyond the scope of this paper, however, based on calculations by \citet{Shull1985}, \citet{Chen2004} proposed 
%-----------
\beal
\label{eq:f_i_ex_Chen}
\tilde{g}^{\rm H}_{\rm ion}(z)=\tilde{g}^{\rm H}_{\rm ex}(z)\approx \frac{1-X_{\rm p}}{3},
\end{align}
%-----------
where $X_{\rm p}=N_{\rm p}/N_{\rm H}$ is the free proton fraction. In the context of cosmological recombination with DM annihilations this approximation has already been used by several authors \citep{Padmanabhan2005, Zhang2006, Mapelli2006, Galli2009, Slatyer2009, Huetsi2009}.
Furthermore, \citet{Padmanabhan2005} also applied a similar expression for the specific ionization and excitation fraction of helium, i.e. $\tilde{g}^{\rm He}_{\rm ion}(z)=\tilde{g}^{\rm He}_{\rm ex}(z)\approx \frac{1-Z_{\ion{He}{ii}}}{3}$, where $Z_{\ion{He}{ii}}=N_{\ion{He}{ii}}/N_{\rm He}$ is the fraction of singly ionized helium atoms relative to the total number of helium nuclei.
In this approximation $g_{\rm h}=[1+2X_{\rm p}+f_{\rm He}(1+2Z_{\ion{He}{ii}})]/3[1+f_{\rm He}]$, where both ionizations and excitations were included.

These are {\it very} rough approximations, since many details of the computations are {\it not represented} or {\it recoverable} in this way. 
For simple order of magnitude computations this approach certainly is acceptable, but for more detailed calculations that aim at including model-dependencies, refinements become necessary. 
For example, without further details it is not easy to say what fraction of excitations are due to photons and which are due to electrons or collisions in more general. 
In the former case, one should introduce modifications to $\tilde{g}^{\rm a}_{\rm ex}(z)$ due to {\it photon escape}, which will strongly depend on the actual density of hydrogen atoms and the expansion rate of the medium, but not only on the ionization fraction. For conditions in our Universe, in that case it will be possible to neglect excitations (e.g. like in \citet{Padmanabhan2005}), since the escape probabilities, $P_{\rm esc}$, in the main resonances of hydrogen and helium are extremely small, so that $\tilde{g}^{\rm a}_{\rm ex}(z)\approx P_{\rm esc} \tilde{g}^{\rm a}_{\rm ex}(z)\approx 0$.
On the other hand, if excitations mainly occur due to interactions with primary or secondary electrons no such modification is necessary. 
At low energies ($E\lesssim 100\,$eV) this seems to be the case \citep{Shull1979}.
As we will see below, the results do depend on this assumption, and a more detailed computation will be necessary.

In addition, \citet{Shull1985} assumed that the ionization fractions of hydrogen and helium are always equal. This assumption is not valid in the cosmological recombination problem, where helium is completely recombined at $z\sim 1700-1800$ while hydrogen is still fully ionized. It is not clear that this extreme case can be obtained from their results, since details in the radiative transfer would be very different, likely leading to non-linear scalings.
Already the computations of \citet{Shull1985} suggest that the scaling of $\tilde{g}^{\rm a}_{\rm i}(z)$ with ionization fraction from neutral to fully ionized media already appears to be {\it faster} than linear.
In the case of hydrogen they provide a fit that is close to
%-----------
\beal
\label{eq:f_i_ion_Shull}
\tilde{g}^{\rm H, Shull}_{\rm ion}(z)\approx \frac{2}{5}\,[1-X^{2/5}_{\rm p}]^{7/4}.
\end{align}
%-----------
Also, their results suggest that for nearly neutral media one has $\tilde{g}^{\rm H}_{\rm ion}\sim 2/5$, $\tilde{g}^{\rm H}_{\rm ex}\sim 1/2$, $\tilde{g}^{\rm He}_{\rm ion}\sim 3/5$ and $\tilde{g}^{\rm He}_{\rm ex}\sim 3/10$ (see their Table 2 with $f_{\rm He}= 10\%$) instead of $\tilde{g}^{\rm H}_{\rm ion}\sim\tilde{g}^{\rm H}_{\rm ex}\sim\tilde{g}^{\rm He}_{\rm ion}\sim\tilde{g}^{\rm He}_{\rm ex}\sim 1/3$.

All this shows, that for more accurate computations within the cosmological context, relatively large differences to the above approximations for $\tilde{g}^{\rm a}_{\rm i}(z)$ can still be expected. A more detailed computation, where we also plan to include doubly ionized helium, excitations of levels with $n>2$, and detailed radiative transfer, will be left for some future work.
However, below we will demonstrate that the cosmological recombination spectrum is sensitive to the form of $\tilde{g}^{\rm a}_{\rm i}(z)$, while for cases in connections with DM annihilations the dependence on $f_{\rm d}$ is much weaker.

\subsection{Modifications to the multi-level recombination code}
\label{sec:rec_code}
%---------------
For the computations presented in this paper we use our multi-level hydrogen and helium recombination code \citep{Chluba2009c}. It in principle allows to include the effect of photons feedback and several other recently considered physical processes that modify the recombination history of the Universe at the percent-level \citep[for overview see][]{Fendt2009, Sunyaev2009}.
However, here we will not include most of these corrections, since the ambiguities introduced due to DM annihilations are much larger in any case. We only want to demonstrate the principle aspects of the problem and show that the cosmological recombination spectrum is sensitive to DM annihilations.

As a first step one should add the term given by Eq.~\eqref{eq:dT_dt} to the normal evolution equation for the temperature of the medium. 
Due to the tight coupling of the photon and electron temperature by Compton scattering, the additional heating will not affect the results of our multi-level recombination code until low redshifts ($z\lesssim 200$), where the energy exchange between electrons and photons becomes inefficient.
However, the {\it continuous heating} of the medium at high redshift in principle will lead to some (small) {\it primordial} $\mu$ or $y$-type spectral distortion of the CMB well before the epoch of recombination \citep[e.g.][]{Zeldovich1969, Sunyaev1970b, Illarionov1975a, Illarionov1975b}. 
We do not take this modification of the background radiation into account, but according to our estimates (see Sect.~\ref{sec:est_y}), for the DM models under discussion here, their annihilation should never lead to any important primordial CMB distortion (e.g. the $y$-parameter was always smaller than $\sim 10^{-10}-10^{-8}$). Therefore the effects discussed in our previous work on pre-recombinational energy release \citep{Chluba2008c} are negligible here.
Note that with respect to those earlier computations, the main difference is that we now include direct ionizations and excitations by DM-induced particles, which, as mentioned in the introduction, can be very efficient.

In the multi-level recombination code, one should also add Eqs.~\eqref{eq:dN_dt_i} and \eqref{eq:dN_dt_ex} to the ground state rate equations of hydrogen and helium, accordingly. 
The additional ionizations introduced by DM annihilations will liberate an electron and (nucleus), and depending on the epoch at which this ionization occurs, the ionization will be directly compensated by the recombination of another electron. In the pre-recombinational epochs one should also allow for direct recombinations to the ground state \citep{Chluba2009c}, while during the recombination epochs electrons will be captured to excited states (with principle quantum numbers $n>2$), potentially liberating several (low frequency) photons in the cascade towards the ground state.
Also, Eq.~\eqref{eq:dN_dt_ex_a} should be subtracted from the rate equation for the \ion{H}{i} 2p state in order to ensure conservation of the electron number. Furthermore, we subtract Eq.~\eqref{eq:dN_dt_ex_b} from the rate equation of the $\HeIlevel{2}{1}{P}{1}$-level, assuming that with $n=2$ only this level is reached. Since the transition rate to the $\HeIlevel{2}{3}{P}{1}$-level is $\sim 10^7$ times smaller, this should be sufficient.
Similarly, we subtract Eqs.~\eqref{eq:dN_dt_i} from the rate equation for the free electrons.

%---------------
\begin{figure}
\centering
\includegraphics[width=\columnwidth]
{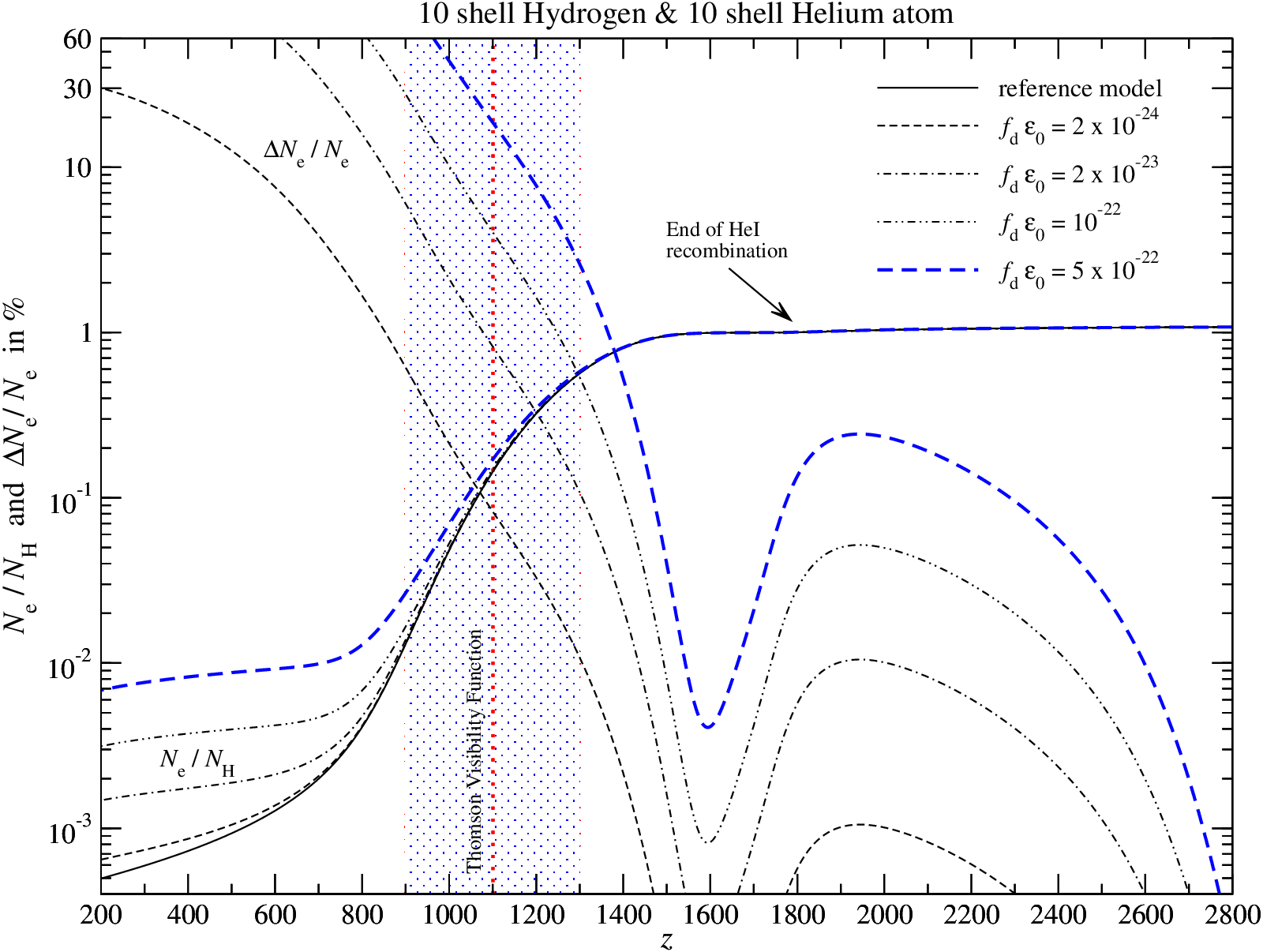}
\caption{Effect of dark matter annihilation on the free electron fraction for different constant values of $f_{\rm d}\epsilon_0$, as labeled. The curves in the upper group show the relative difference in the free electron fraction in comparison to the reference model \citep{Jose2008}. The shaded area indicates the region around the Thomson visibility function, which defines the last scattering surface.}
\label{fig:Ne.DNe_Ne}
\end{figure}
%---------------

%---------------
\begin{figure}
\centering
\includegraphics[width=\columnwidth]
{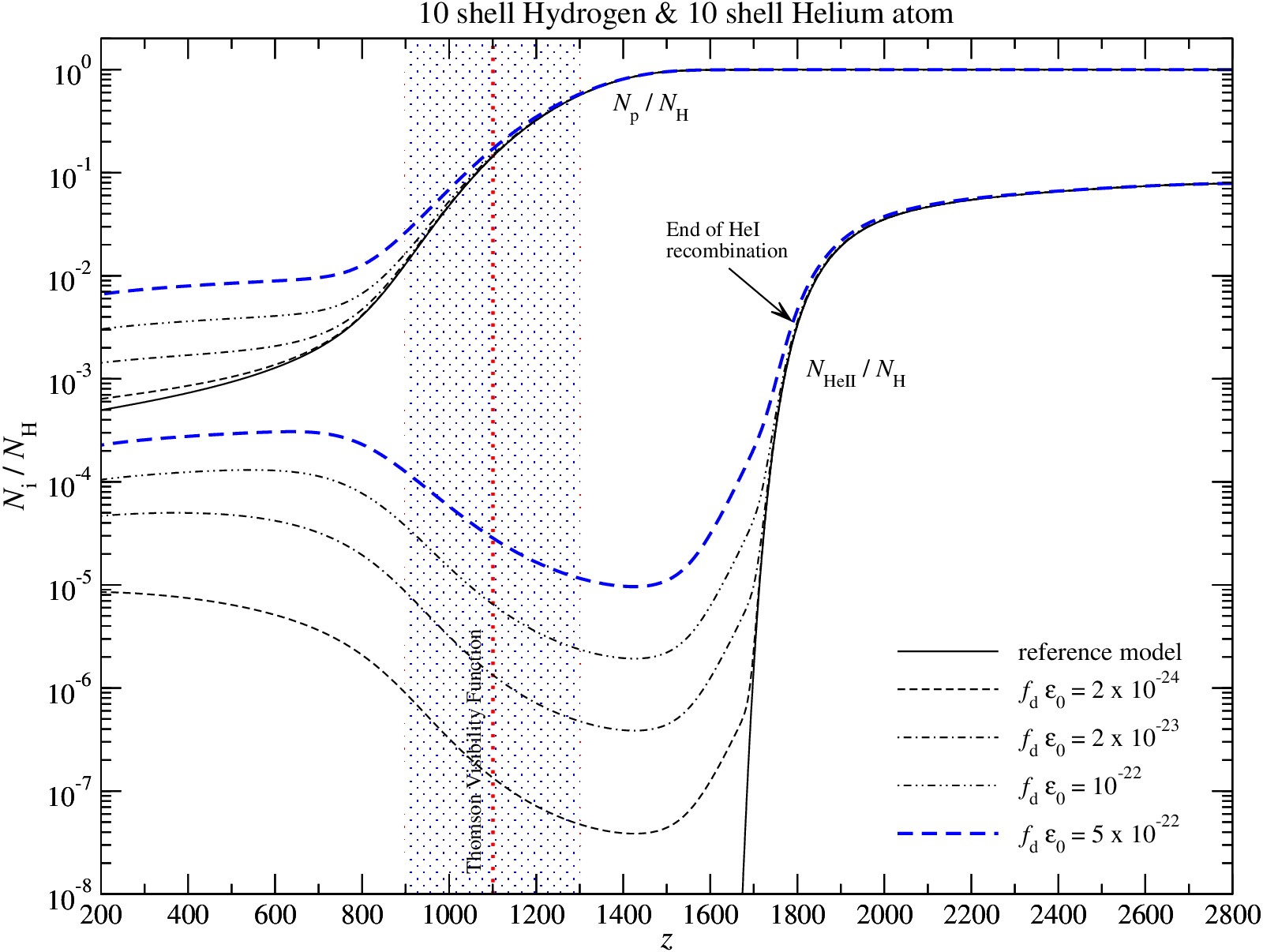}
\caption{Effect of dark matter annihilation on the number density of free protons, $N_{\rm p}$, and $\ion{He}{ii}$ ions, $N_\ion{He}{ii}$, for different constant values of $f_{\rm d}\epsilon_0$, as labeled. The shaded area indicates the region around the Thomson visibility function, which defines the last scattering surface. For comparison we also show the result for our reference model \citep{Jose2008}.}
\label{fig:Np.NHeII}
\end{figure}
%---------------

%++++++++++++++++++++++++++++++++++++++
%
%
%
%++++++++++++++++++++++++++++++++++++++
\section{How does the cosmological recombination spectrum change due to dark matter annihilations?}
\label{sec:DM_ann_dependencies}
%---------------
In this section we illustrate the dependencies of the cosmological recombination spectrum and ionization history on the energy injection by annihilating DM.
We start the discussion with cases that assume $f_{\rm d}(z)\,{\epsilon}_0={\rm const}$, and then go to more complicated models in the following. In particular we want to demonstrate that the cosmological recombination spectrum is sensitive to differences in the DM annihilation model, in particular related to the branching of the deposited energy into heating, ionizations and excitations.

\subsection{Effect of DM annihilation on $N_{\rm e}$: constant $f_{\rm d}(z)\,{\epsilon}_0$}
\label{sec:DM_ann_Ne_const_feps}
%---------------
Figure \ref{fig:Ne.DNe_Ne} shows the effect of DM annihilation on the free electron fraction. 
We included 10 shells for hydrogen and helium into our computations and assumed that all $\tilde{g}^{\rm a}_{\rm i}(z)$ are given by Eq.~\eqref{eq:f_i_ex_Chen}.
One can clearly see that DM annihilations have the strongest effect at low redshifts, leading to a delay of recombination (at $z\sim 1000$) and an increase in the residual free electron fraction at very low redshift ($z\sim 200$). The recombination of neutral helium (at $z\sim 2000$) is hardly changing, even in the most extreme cases considered here, implying that the net recombination rate for helium is not affected as strongly by ionizations due to DM annihilation.

One reason for this behavior is that for a given energy deposition rate, $\id E_{\rm d}/\id t$, due to the difference in the ionization potentials there are about $\sim 2$ times fewer ionizing and photons per helium atom available than for hydrogen (see Eq.~\eqref{eq:dN_dt_i} for confirmation). 
Similarly, the effective DM-induced excitation rate is $\sim 2$ times smaller.
%
%However, in comparison to hydrogen this should not lead to a very large difference in the effect of DM annihilation on the helium recombination history.
%
Another reason is that any small relative difference $\Delta N^{\rm He}_{\rm e}/N^{\rm He}_{\rm e}$ in the number of free electrons from helium, due to its small abundance ($\sim 8\%$ in comparison to hydrogen), will have a $\sim 13$ times smaller effect on the total ionization history $N_{\rm e}=N^{\rm H}_{\rm e}+N^{\rm He}_{\rm e}$, which includes the electrons from hydrogen.
In the early stages of helium recombination one therefore expects that the ionization history can only be affected by a comparable amount as during hydrogen recombination when increasing the DM annihilation rate $\sim 20-30$ times. Looking at Fig.~\ref{fig:Ne.DNe_Ne}, and comparing the curves for $f_{\rm d}(z)\,{\epsilon}_0=\pot{2}{-23}$ and $f_{\rm d}(z)\,{\epsilon}_0=\pot{5}{-22}$ at $z\sim 1100$ and $z\sim 2000$, seems to confirm this statement.

However, towards the end of helium recombination the main reason for the rather small effect of DM annihilations on the free electron fraction is connected with the acceleration of helium recombination caused by the absorption of resonant \ion{He}{i} photons in the Lyman continuum of hydrogen \citep{Switzer2007I, Kholupenko2007, Jose2008}.
This shifts the end of helium recombination from $z\sim 1600$ to $z\sim 1750$, because the effective recombination rate of helium is increased many times by this processes. 
It is extremely hard to delay helium recombination with DM annihilations, once this process is working well ($z\lesssim 1900$). 
Here it is also important that the \ion{He}{ii} ions interact with a {\it bath} of free electrons from hydrogen. 
Per \ion{He}{ii} ion there are about $13$ electrons available for recombinations, while per proton during hydrogen recombination there is only one. This number of electrons per \ion{He}{ii} ion remains practically constant until the recombination of hydrogen begins. 
Therefore, without DM annihilations practically all helium atoms recombine\footnote{This would also be true without the inclusion of the speed up due to the \ion{H}{i} continuum opacity.}, leaving basically no free \ion{He}{ii} ions at low redshifts (see Fig.~\ref{fig:Np.NHeII}).

Only at very late stages, when recombinations of helium are already slow, one again expects some modifications in the helium ionization history. This is because there DM annihilations can (partially) reionize helium atoms, without this process being (significantly) reversed by recombinations. 
In Fig.~\ref{fig:Np.NHeII} one can see that at low redshifts the number of residual \ion{He}{ii} ions indeed increases strongly when accounting for DM annihilations. Nevertheless, in all considered cases no more than $\sim 0.3\%$ of helium atoms are reionized by DM annihilation at low redshifts, and the total contribution of free electrons from helium to the residual free electron fraction at $z\sim 200$ does not exceed a few percent.

%\subsubsection{Effect on the CMB power spectra}
%\label{sec:Ne_Cl}
%%---------------

%---------------
\begin{figure}
\centering
\includegraphics[width=\columnwidth]
{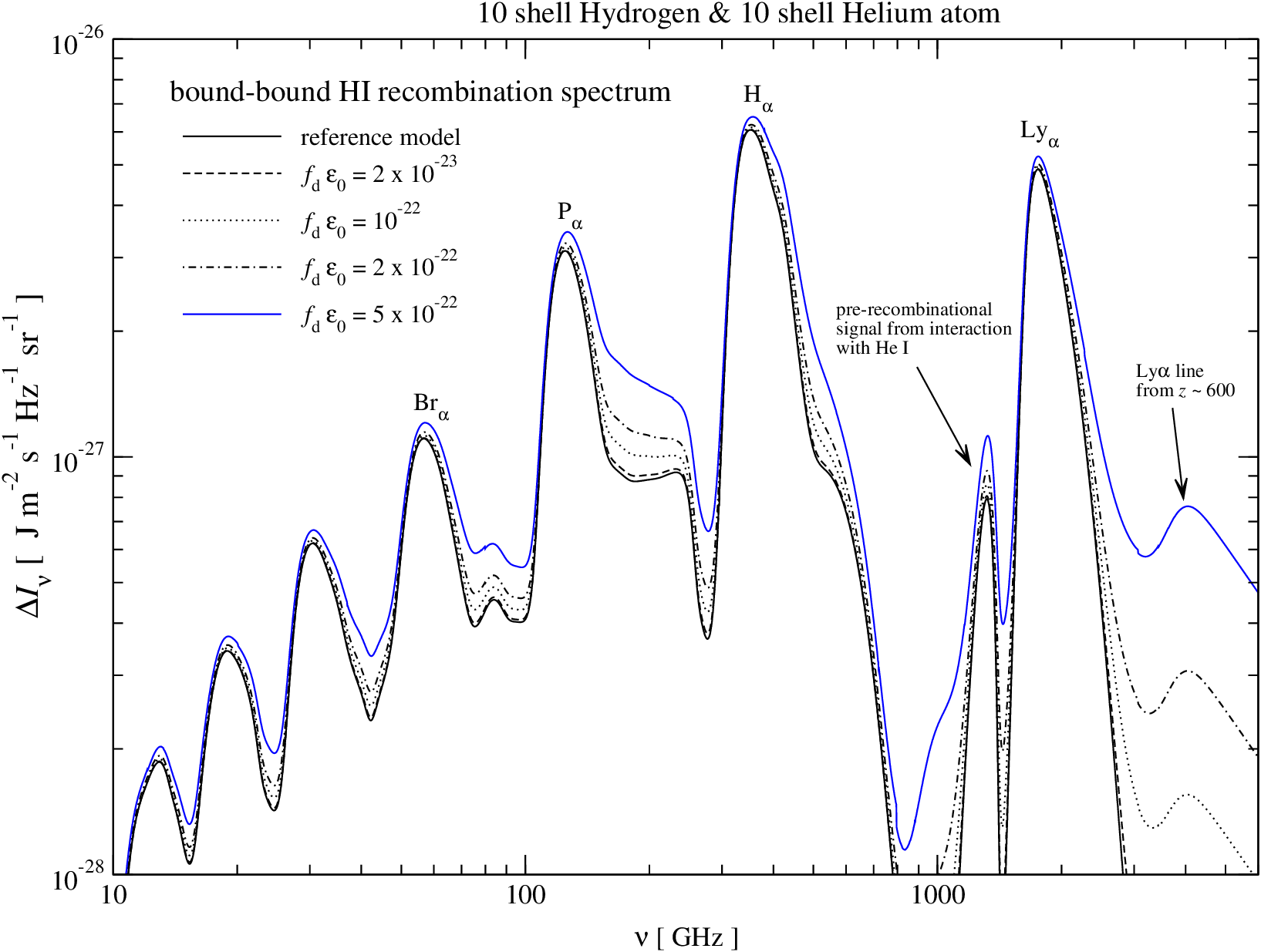}
\\[2mm]
\includegraphics[width=\columnwidth]
{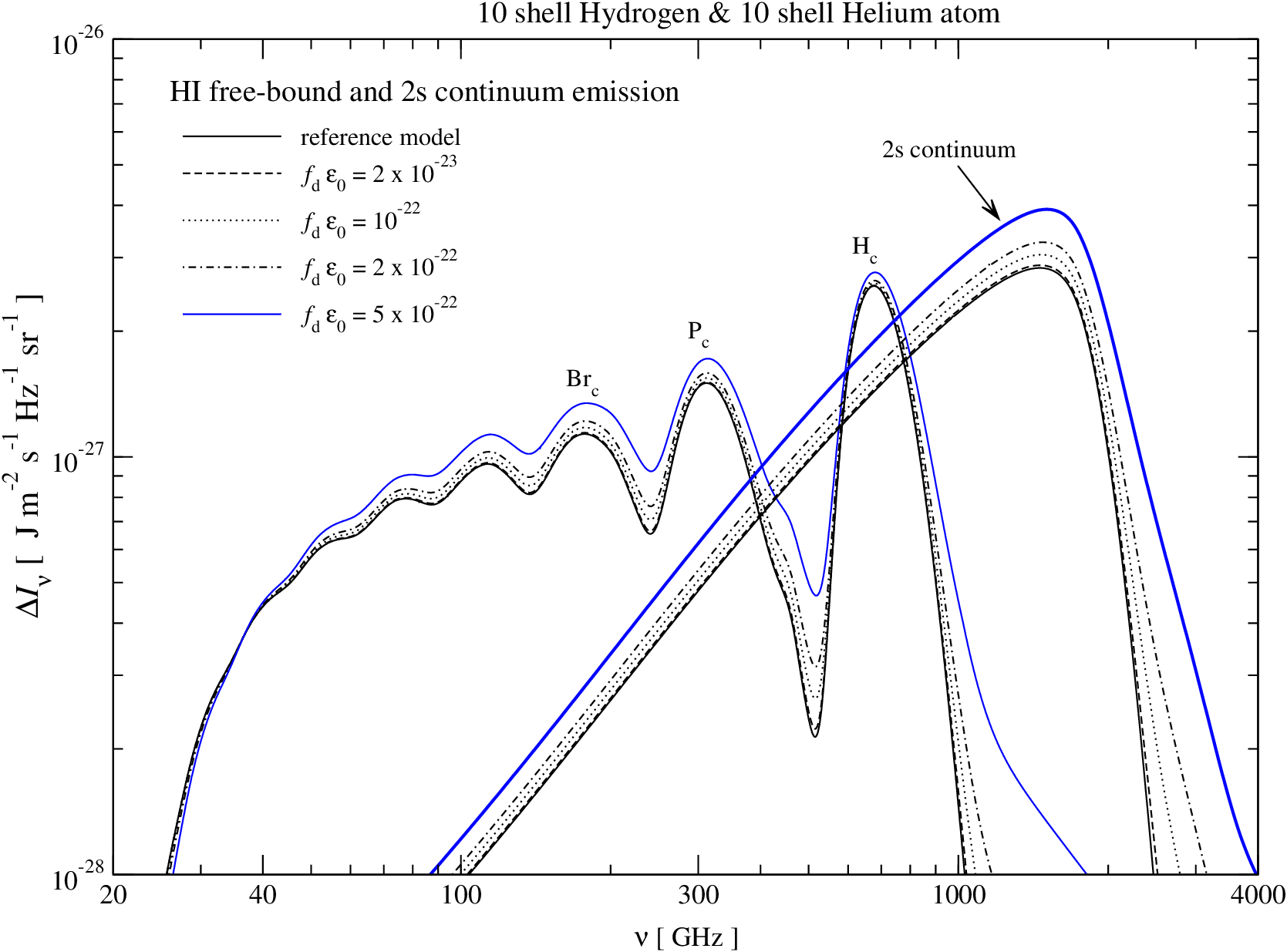}
\caption{Effect of dark matter annihilation on the \ion{H}{i} bound-bound dipole recombination spectrum (upper panel), and free-bound and 2s-1s two-photon continuum emission (lower panel) for different constant values of $f_{\rm d}\epsilon_0$, as labeled. Shown are the {\it present day} ($z=0$) spectral distortion of the CMB generated by hydrogen during cosmological recombination including 10 shells in the computations.
For comparison we also show the result for our reference model \citep{Jose2008}. }
\label{fig:spec.HI.A}
\end{figure}
%---------------

\subsection{Effect of DM annihilation on the cosmological recombination spectrum: constant $f_{\rm d}(z)\,{\epsilon}_0$}
\label{sec:DM_ann_Spec_const_feps}
%---------------
As we have seen in the previous section, the largest modifications in the free electron fraction appear at the end of hydrogen recombination. For the Thomson visibility function and the CMB power spectra the modifications around $z\sim 1100$ are most important, while the {\it huge} relative changes in the residual electron fraction at $z\lesssim 500$ actually do not matter that much.

It is known that the recombination lines from hydrogen mainly appear at $z\sim1300-1400$ \citep[e.g. see][]{Jose2006, Chluba2006b, Chluba2007}, where about $\sim20\%$ of the hydrogen atoms recombined, while at maximum visibility  ($z\sim 1100$) already $\sim86\%$ of all \ion{H}{i} was formed.
From the differences in the free electron fraction it is therefore already clear that the \ion{H}{i} recombination lines will mainly be modified on the {\it blue} sides of the recombination features, with an increase of the emission due to additional ionizations and subsequent recombinations. With increasing DM annihilation efficiency the changes will become more strong and should eventually also affect the maxima of the recombination features. This will lead to {\it shifts} in their {\it positions} towards higher frequencies, and an {\it increase} in the overall {\it amplitude} and {\it width} of the recombination features.

In Fig.~\ref{fig:spec.HI.A} we illustrate this behavior of the different component in the \ion{H}{i} recombination spectrum when taking the effect of DM annihilations into account.
We included 10 shells for hydrogen and helium into our computations and assumed that all $\tilde{g}^{\rm a}_{\rm i}(z)$ are given by Eq.~\eqref{eq:f_i_ex_Chen}.
For all components the DM annihilation increases the overall amplitude of the emission and hence the total number of photons released during recombination.
This is simply related to the fact that every {\it extra} ionization caused by DM annihilation will liberate an electron which then can recombine to some excited state of hydrogen. From there it will cascade towards lower levels emitting several photons on its way. These photons are released in addition to those from the normal recombination epoch. 
Because that at $z\lesssim 2000$ the \ion{H}{i} Lyman continuum is completely blocked \citep[e.g.][]{Chluba2007b}, every ionization caused by DM annihilations will lead to {\it at least} two photons (i.e. Balmer continuum and Lyman $\alpha$) at lower frequency.
However, since electrons also can be captured into some highly excited state ($n>2$), the effective number of emitted photons per ionization can be larger than two.
In Fig.~\ref{fig:spec.HI.A} one can see that indeed also the emission in transition among highly excited levels, appearing at low frequencies in the cosmological recombination spectrum, increases when including the effect of DM annihilations. This shows that a significant number of electrons are captured to states with $n>2$.

Furthermore, from Fig.~\ref{fig:spec.HI.A} it is clear that in particular for the \ion{H}{i} bound-bound dipole emission lines the positions and widths of the recombination features are affected by the DM annihilations.
Both the free-bound and 2s-1s two-photon continuum emission are less sensitive in this respect, since they are initially very broad.
In this context, especially the spectral features due to Paschen $\alpha$ (visible at $\nu \sim 120\,$GHz) and Balmer $\alpha$ (visible at $\nu \sim 350\,$GHz) are interesting, as they are both very prominent and not overlapping so much with other spectral features. 
Nevertheless, the distortions from the recombination epoch are affected in basically {\it all} spectral bands. From an observational point of view it will be important to look at the CMB distortions in many frequency channels and to determine the positions and width of several features simultaneously.

Here we would also like to mention that the very high frequency spectral feature visible at $\nu\sim 4\,$THz is created by the \ion{H}{i} Lyman $\alpha$ resonance at $z\sim 600$. Although from an observational point of view this distortion is not very interesting (the cosmic infrared background is far too strong in this spectral band), for computations of the low redshift {\it chemistry} \citep[e.g. see][for recent computations]{Schleicher2008}, also including the effect of non-equilibrium background radiation \citep{Switzer2005, Vonlanthen2009}, such feature may be relevant. However, in this case processes directly related to secondary particles from DM annihilation may still be more important \citep[e.g. like for example found in the case of cosmic rays][]{Jasche2007}.

%---------------
\begin{figure}
\centering
\includegraphics[width=\columnwidth]
{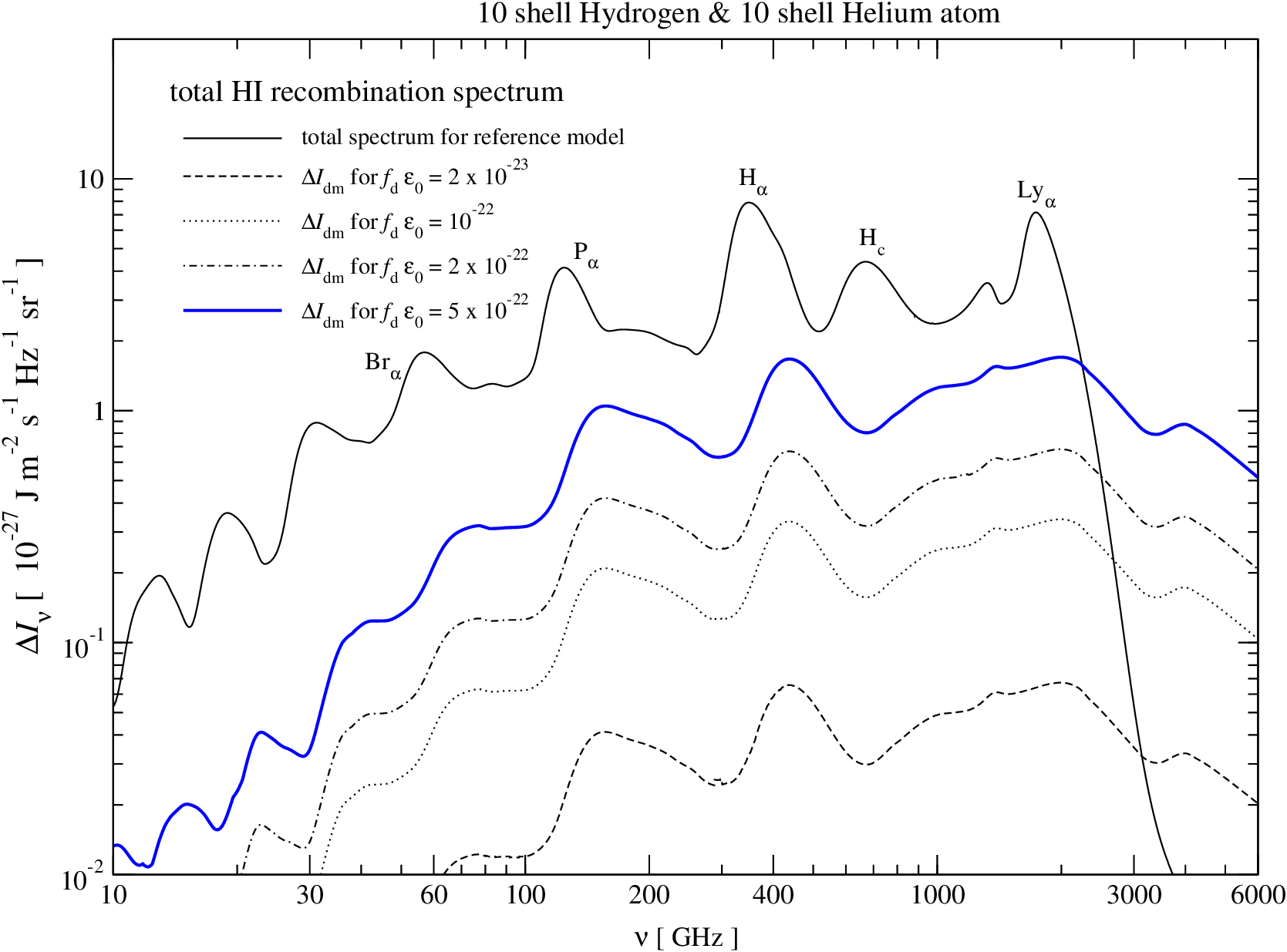}
\caption{Component to the total \ion{H}{i} cosmological recombination spectrum which is induced by dark matter annihilation for different constant values of $f_{\rm d}\epsilon_0$, as labeled. 
We computed the absolute difference, $\Delta I_{\rm dm}(\nu)$, in the {\it total}, present-day \ion{H}{i} recombination spectrum with respect to our reference model \citep{Jose2008} which does not include DM annihilations (shown as thin solid line).
}
\label{fig:DI_dm.HI}
\end{figure}
%---------------
\subsubsection{The \ion{H}{i} spectral distortion produced by DM annihilations}
\label{sec:DM_ann_DI_dm}
%---------------
For Fig.~\ref{fig:DI_dm.HI} we also separated the contribution, $\Delta I_{\rm dm}(\nu)$, to the total \ion{H}{i} cosmological recombination spectrum (bound-bound $+$ free-bound $+$ 2s-1s two-photon continuum) which is induced by dark matter annihilations.
The main broad features at $\nu\sim 440\,$GHz, $150\,$GHz and $70\,$GHz in $\Delta I_{\rm dm}(\nu)$ are due to {\it reactivation} of the Balmer $\alpha$, Paschen $\alpha$ and Brackett $\alpha$ resonance at $z\lesssim 1300-1400$, respectively.
The width of the associated spectral features is comparable to those produced in the normal recombination process. 
Furthermore, one can see that $\Delta I_{\rm dm}(\nu)\propto f_{\rm d}\epsilon_0$, so that it is possible to compute the total \ion{H}{i} recombination spectrum by adding this component (with appropriate rescaling) to the normal recombination spectrum.
For $f_{\rm d}\epsilon_0\sim 10^{-22}$ the relative contribution of the DM-induced emission on average reaches a few percent, and in some bands up to $\sim 10\%$.
In the most extreme case presented here the relative change in the recombination spectrum introduced by DM annihilations reaches $\sim 60\%-70\%$ in the vicinity of the Balmer and Paschen series.

{\it Why does $\Delta I_{\rm dm}(\nu)$ actually exhibit spectral features?} 
For this it is important that although DM particles continuously annihilate, the fraction of energy $\tilde{g}^{\rm H}_{\rm ion}(z)\propto [1-X_{\rm p}]$ that actually goes into ionizations is very small until the recombination epoch of hydrogen is entered.
Similarly, excitations of neutral atoms caused by DM annihilations are not effective until hydrogen starts recombining. 
On the other hand, the energy deposition rate per hydrogen atom scales like $\frac{1}{N_{\rm H}}\frac{\id E_{\rm d}}{\id t}\propto f_{\rm d}(z)[1+z]^3$, and hence decreases with redshift. 
Therefore, the product $f_{\rm d}(z)[1+z]^3\,\tilde{g}^{\rm H}_{\rm ion}(z)$ exhibits a maximum, that is mainly determined by $\tilde{g}^{\rm H}_{\rm ion}(z)$ which changes rather fast close to the recombination epoch.
As a consequence, the spectral distortions which are introduced by DM annihilations also show maxima with a width which is comparable to the duration of recombination. 
As we will see below (Sect.~\ref{sec:DM_ann_Spec_fd}), the DM-induced changes to the \ion{H}{i} recombination spectrum are indeed sensitive to the redshift dependence of $\tilde{g}^{\rm H}_{\rm ion}(z)$ and to a smaller extend of $f_{\rm d}(z)$.

%---------------
\begin{figure}
\centering
\includegraphics[width=\columnwidth]
{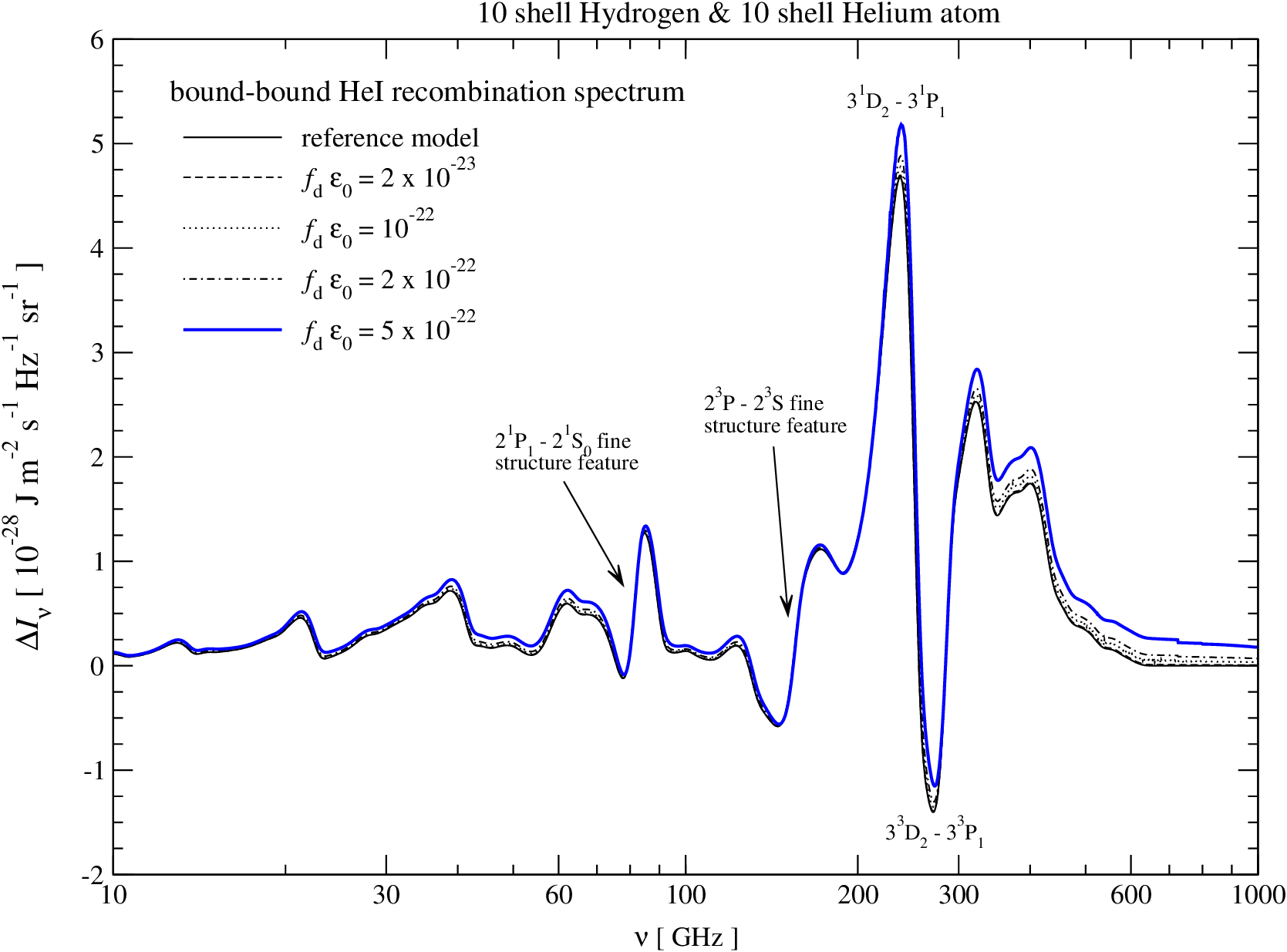}
\caption{Effect of dark matter annihilation on the \ion{He}{i} bound-bound dipole recombination spectrum for different constant values of $f_{\rm d}\epsilon_0$, as labeled. Shown are the {\it present day} ($z=0$) spectral distortion of the CMB generated by neutral helium during cosmological recombination including 10 shells in the computations.
For comparison we also show the result for our reference model \citep{Jose2008}. }
\label{fig:spec.HeI.A}
\end{figure}
%---------------
\subsubsection{Effect on the helium recombination spectrum}
\label{sec:DM_ann_HeI_Spec_const_feps_ex}
%---------------
In Fig.~\ref{fig:spec.HeI.A} we also present the changes in the CMB spectral distortions introduced by \ion{He}{i}. Here the effect of DM annihilations is significantly smaller than for hydrogen: for $f_{\rm d}\epsilon_0=\pot{5}{-22}$ the differences are of the order of $\sim 10\%-20\%$ in some bands, while for hydrogen changes up to $\sim 60\%-70\%$ were found (cf. upper panel in Fig.~\ref{fig:spec.HI.A}).
Still this is much larger than the changes seen in the helium ionization history, where for $f_{\rm d}\epsilon_0=\pot{5}{-22}$ the corrections were $\Delta N^{\rm He}_{\rm e}/N^{\rm He}_{\rm e}\sim 0.1\%\times 13\sim 1\%$ (cf. Fig.~\ref{fig:Ne.DNe_Ne}).
This implies that the recombination spectrum is more sensitive to energy deposition during helium recombination than the ionization history itself.
Also it is clear that the small changes in $N_{\rm e}$ during helium recombination will not propagate very much to the CMB power spectra, so that one cannot expect to see any signature of DM annihilation during helium recombination in the $C_l$'s. However, directly observing the helium recombination spectrum in principle could shed light on processes occurring during this epoch. %, even though the changes are very small.

{\it How does this work?}
In the early stages of helium recombination and in its pre-recombinational epoch ($z\gtrsim 2600$) the number of extra \ion{He}{i} ionizations caused by DM annihilations is very small ($\tilde{g}^{\rm He}_{\rm ion}(z)\propto [1-Z_{\ion{He}{ii}}]\ll 1$). According to our parametrization most of the deposited energy is going into heating of the medium at that time (Sect.~\ref{sec:ion_ex}).
The same is true for the pre-recombinational epoch of hydrogen, explaining why practically no extra emission is produced, even though DM is continuously annihilating (Sect.~\ref{sec:DM_ann_DI_dm}).

Our discussion in Sect.~\ref{sec:DM_ann_Ne_const_feps} has already shown that the ionization history of helium is not affected as much by DM annihilations, so that from this very few extra photons are expected. 
There it was most important that the small fraction of neutral hydrogen present at  the end of helium recombination leads to a {\it huge} increase in the photons escape probability, so that helium recombination is strongly accelerated.
This makes it very hard to change the \ion{He}{i} ionization fraction by DM annihilations, since practically every extra ionization is {\it directly} reversed by a recombination.

However, during the {\it whole} epoch of helium recombination, DM annihilations do lead to some {\it extra} ionizations. 
This drives {\it loops} of transitions, that start with the ionization of a neutral helium atom by DM annihilation, and end with the release of extra photons at lower frequencies in the cascade of electrons from excited levels towards the ground state.
%until the amount of free \ion{He}{ii} ions has become too small that recombinations after ionizations become very unlikely. 
%
This explains why it is possible to see the effect of extra ionizations in the helium recombination spectrum, while at the same time the effect on the ionization history is much smaller. 
However, due to the small abundance of helium in our Universe these extra photons are not as important for the total recombination radiation as those from hydrogen. Still for accurate computations of a {\it spectral template} one should also take these into account, but in this case also aspects related to details in the Lyman $\alpha$ radiative transfer \citep{Chluba2009, Chluba2009b}, feedback \citep{Chluba2009c}, and electron scattering \citep{Jose2008, Chluba2008d} will become important.

In addition, one should mention, that, for example, in the case of decaying particles it is possible that during helium recombination much more energy is released than during hydrogen recombination. In that situation the total emission from helium could be increased many times, although the changes to the hydrogen recombination spectrum may still be small. Similarly, the modifications to the ionization history might still be not important for computations of the CMB power spectra, so that only the cosmological recombination spectrum will allow us to put constraints on possible extra energy release in this case (see Sect.~\ref{sec:disc} for more discussion).

\begin{figure}
\centering
\includegraphics[width=\columnwidth]
{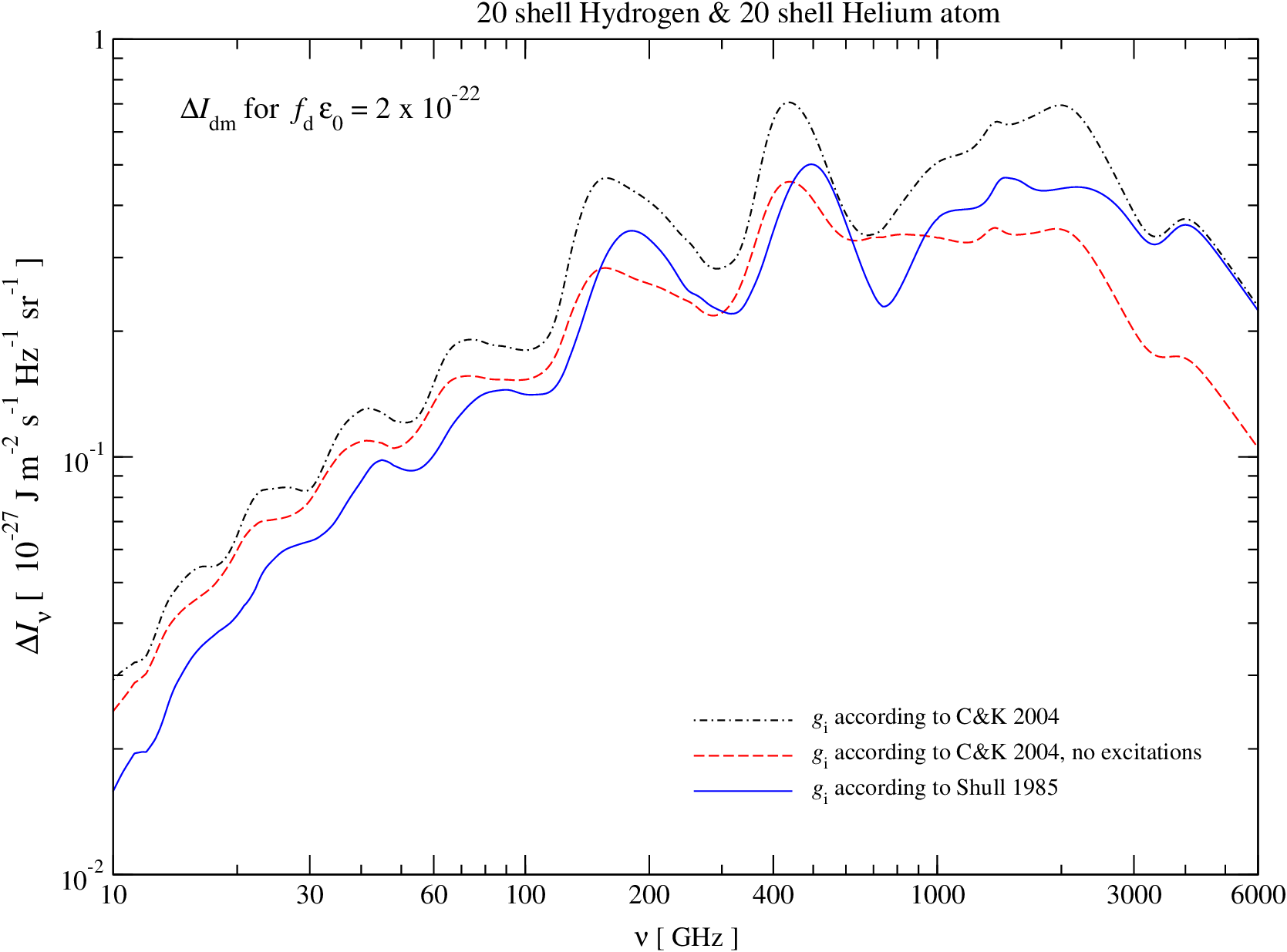}
\caption{Contribution to the total \ion{H}{i} cosmological recombination spectrum which is induced by dark matter annihilation for $f_{\rm d}\epsilon_0=\pot{2}{-22}$ and different choices of $\tilde{g}^{\rm a}_{\rm i}$.
We included 20 shells for hydrogen and helium in our computations.
Like in Fig.~\ref{fig:DI_dm.HI} we give  the absolute difference, $\Delta I_{\rm dm}(\nu)$, in the {\it total}, present-day \ion{H}{i} recombination spectrum with respect to our reference model \citep{Jose2008} which does not include DM annihilations.
For the dashed-dotted line we used $\tilde{g}^{\rm a}_{\rm i}$ as given by Eq.~\eqref{eq:f_i_ex_Chen}, for both hydrogen and helium. In the case 'no excitations' we excluded the DM-induced excitations of hydrogen and helium. 
For the solid curve we use the expression~\eqref{eq:f_i_ion_Shull} for all $\tilde{g}^{\rm a}_{\rm i}$.
}
\label{fig:DI_dm.HI.ex}
\end{figure}
%---------------

%---------------
\begin{figure}
\centering
\includegraphics[width=\columnwidth]
{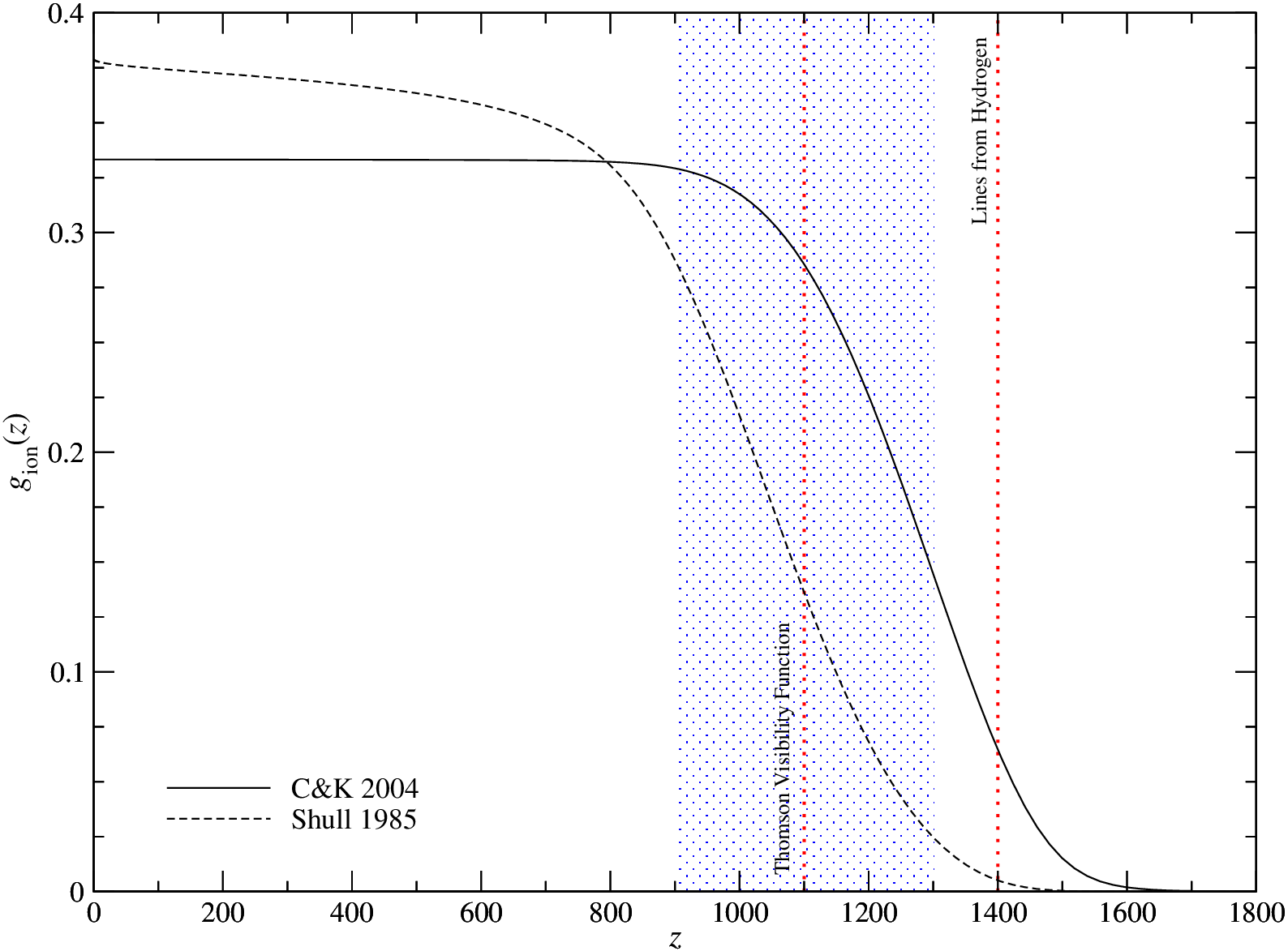}
\caption{Ionization efficiency, $\tilde{g}^{\rm H}_{\rm d}$, for the standard ionization history. The solid like corresponds to the approximation proposed by \citet{Chen2004}, while the dashed curve is motivated by the work of \citet{Shull1985}. We also marked the region around the last scattering surface ($z\sim 1100$), and where most of the recombination lines from hydrogen ($z\sim 1400$) are produced.}
\label{fig:gt}
\end{figure}
%---------------
\subsection{Dependence of the cosmological recombination spectrum on ionization and excitation efficiencies $\tilde{g}^{\rm a}_{\rm i}$}
\label{sec:DM_ann_Spec_const_feps_g}
%---------------
In the previous sections we have assumed that all the $\tilde{g}^{\rm a}_{\rm i}$ are given by the expression~\eqref{eq:f_i_ex_Chen}, which was initially suggested by \citet{Chen2004}. 
In this section we demonstrate that the DM-induced contribution to the cosmological recombination spectrum is sensitive to the branching of the deposited energy into heating, ionization and excitations.
This emphasizes how important it is to refine the modelling of the effect of DM annihilations on the recombination process, especially when aiming at computing detailed templates for the recombination spectrum or accurately accounting for this process in connection with the CMB power spectra.

\subsubsection{Changes in the recombination spectrum when excitations by DM annihilations are not included}
\label{sec:DM_ann_Spec_const_feps_no_ex}
%---------------
As a first case, we will assume that excitations caused by DM annihilations are negligible. If additional excitations are mediated by photons, such approximation will be more appropriate.
It is clear that this will {\it reduce} the amount of extra low frequency emission, since due to the strong coupling of electrons in the 2p state to the 3d and 3s state, and the continuum \citep{Chluba2008b, Chluba2009}, every additional excitation to the second shell also leads to some additional ionizations and transitions among highly excited states.

In Fig.~\ref{fig:DI_dm.HI.ex} we show the DM-induced contribution to the total \ion{H}{i} cosmological recombination spectrum when neglecting excitations by DM annihilations. 
At low frequencies the overall amplitude of the distortion is reduced by $\sim 15\%-20\%$, while at high frequencies the distortions are about two times lower  than in the case which includes excitations.
Also the changes close to the Balmer and Paschen features reduced by a factor of $\sim 1.5$.
For the same $f_{\rm d}\epsilon_0$ one therefore finds a smaller admixture of the DM-induced signal to the normal cosmological recombination spectrum. This implies that the changes in the width and position of the recombination lines will be smaller when excitations are not efficient. 

It is clear that part of the difference can be compensated by increasing the effective value $f_{\rm d}\epsilon_0$, but since also the relative amplitudes of the features are affected (e.g. low to high frequency contributions), a differential signal remains. This in principle should allow to determine how efficient excitations from DM annihilations are. As mentioned in Sect.~\ref{sec:f_ion_ex} this will depend on details of the DM model and the annihilation channels that are important.

\subsubsection{Direct dependence on the redshift scaling of $\tilde{g}^{\rm a}_{\rm i}$}
\label{sec:DM_ann_Spec_const_feps_Shull}
%---------------
As pointed out by \citet{Chen2004}, and as also mentioned in Sect.~\ref{sec:f_ion_ex}, the approximation~\eqref{eq:f_i_ex_Chen} is very crude and does not capture most of the real dependencies of the branching of the deposited energy into heating, excitations and ionizations on the ionization degree, density of the plasma, helium abundance and expansion rate of the Universe. 
However, the modifications to the recombination spectrum do depend on the scaling of $\tilde{g}^{\rm a}_{\rm i}$ with redshift.
As another example, we therefore also compute the changes in the recombination spectrum using the approximation~\eqref{eq:f_i_ion_Shull} instead of \eqref{eq:f_i_ex_Chen}.

In Fig.~\ref{fig:gt} we present the comparison of these two approximations for the standard ionization history computed with {\sc Recfast} \citep{SeagerRecfast1999}.
As one can see, the ionization efficiency rises much slower when using the expression~\eqref{eq:f_i_ion_Shull}, which is more closely based on the work of  \citet{Shull1985} than Eq.~\eqref{eq:f_i_ex_Chen}.
At $z\sim 1100$, i.e close to the maximum of the Thomson visibility function, $\tilde{g}^{\rm H, Shull}_{\rm ion}$ is practically two times smaller than $\tilde{g}^{\rm H, Chen}_{\rm ion}$.
This implies that ionizations will become important significantly later, implying that also the maxima of the DM-induced spectral features should be shifted towards higher frequencies (see Sect.~\ref{sec:DM_ann_DI_dm} for explanation).
Furthermore one expects that the overall amplitude of the additional distortions should be smaller, since the energy deposition rate $\id E_{\rm d}/\id t$ decreases with redshift.

%---------------
\begin{figure}
\centering
\includegraphics[width=\columnwidth]
{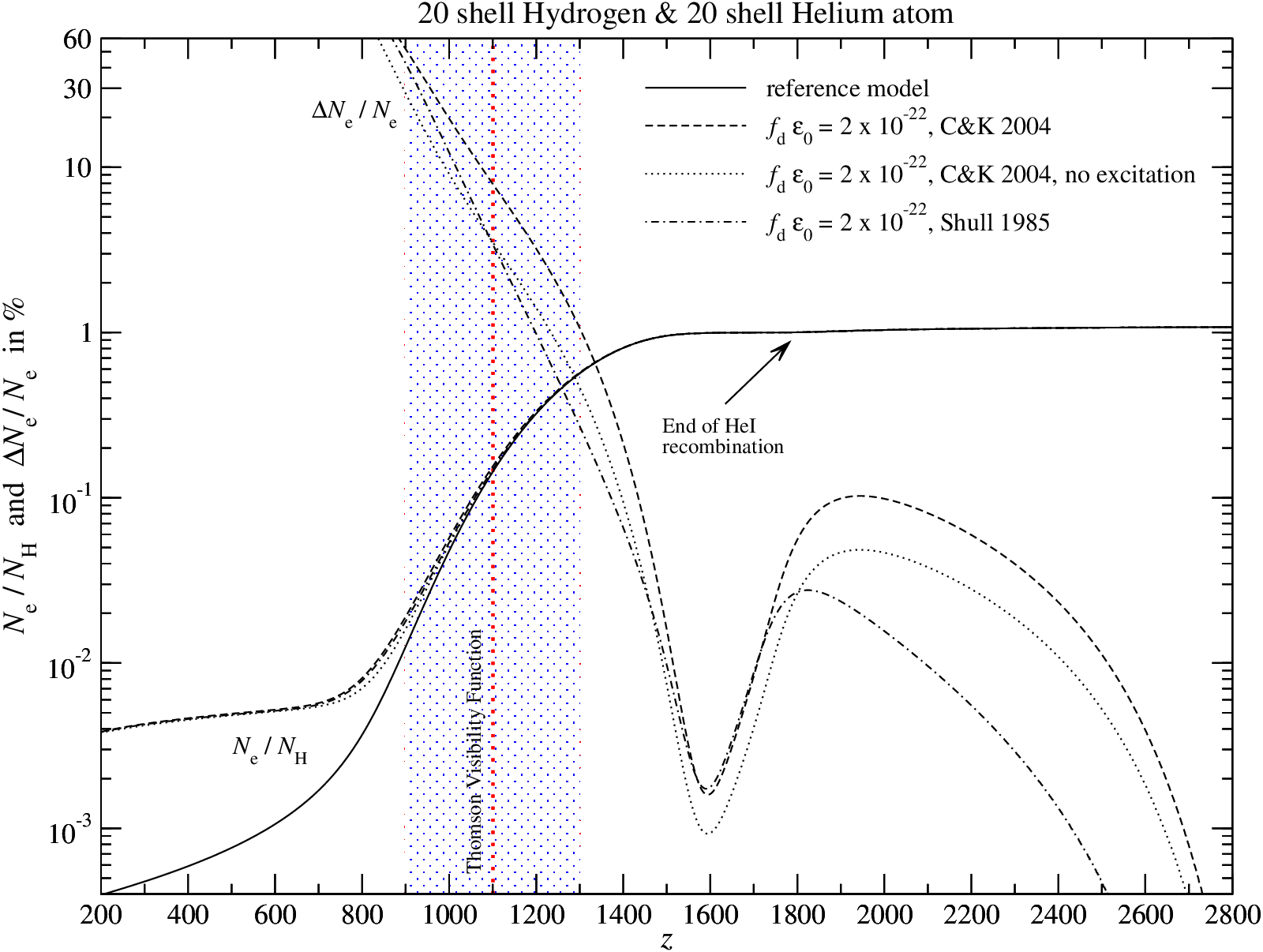}
\caption{Effect of dark matter annihilation on the free electron fraction for $f_{\rm d}\epsilon_0=\pot{2}{-22}$ and different choices of $\tilde{g}^{\rm a}_{\rm i}$.
We included 20 shells for hydrogen and helium in our computations.
The curves in the upper group show the relative difference in the free electron fraction in comparison to the reference model \citep{Jose2008}. The shaded area indicates the region around the Thomson visibility function, which defines the last scattering surface.
For the dashed-dotted line we used $\tilde{g}^{\rm a}_{\rm i}$ as given by Eq.~\eqref{eq:f_i_ex_Chen}, for both hydrogen and helium. In the case 'no excitations' we excluded the DM-induced excitations of hydrogen and helium. 
For the solid curve we use the expression~\eqref{eq:f_i_ion_Shull} for all $\tilde{g}^{\rm a}_{\rm i}$.}
\label{fig:Ne.DNe_Ne.20.gex}
\end{figure}
%---------------
%
In Fig.~\ref{fig:DI_dm.HI.ex} we present the DM-induced distortion when using $\tilde{g}^{\rm H, Shull}_{\rm ion}$ for all $\tilde{g}^{\rm a}_{\rm i}$. Indeed the  maxima of the distortions are shifted by $\Delta\nu/\nu\sim 10\%-20\%$ and the overall amplitude reduced by $20\%-50\%$ at different frequencies. 
This shows that the cosmological recombination spectrum is rather sensitive to the detailed time-dependence of the ionization and excitation efficiencies.
Given the large uncertainty in these functions in the current computations, it will be important to refine the modelling of the energy deposition by DM annihilations in this respect.
This will also be important in connection with precise computations of the ionization history and CMB power spectra, and the obtained limits on models of annihilating DM using current CMB data \citep[e.g.][]{Galli2009, Slatyer2009, Huetsi2009}.

\subsection{Effect of DM annihilations including the time-dependence of $f_{\rm d}$}
\label{sec:DM_ann_Spec_fd}
%---------------
In the previous sections we have seen that the {\it shape} of the additional spectral distortion did not change very much when increasing the DM annihilation efficiency (see Fig.~\ref{fig:DI_dm.HI}). 
However, neglecting the effect of excitations or changing the redshift dependence of $\tilde{g}^{\rm a}_{\rm i}$ did lead to some notable modifications in the shape of the distortions (cf. Fig.~\ref{fig:DI_dm.HI.ex}).
Similarly, one expects that including the additional {\it time-dependence} in the energy deposition rate will affect the distortions.
Here we want to compare the distortions for different models of annihilating DM, however, as we show below the model-dependence of $f_{\rm d}$ introduces only rather small differences.

Inspecting the results of \citet{Slatyer2009} for the energy deposition efficiencies, $f_{\rm d}(z)$, for different models of annihilating DM, one can see that in most cases {\it only} the overall amplitude of $f_{\rm d}$ is changing, while the shape is very similar, resembling the one for $\chi\bar{\chi}\rightarrow e^+e^-$ DM annihilation with $M_{\chi}=100\,$GeV. 
This leads to a strong {\it degeneracy}, since such changes in the overall amplitude can be {\it compensated} when allowing for appropriate (constant) boost factors to the annihilation cross-section, e.g. motivated by the effect of Sommerfeld enhancement. Therefore, we expect that in all these cases the shape of the DM-induced spectral distortion will be very similar, and that only the amplitude will depend on the specific model via an overall efficiency factor.
%

%---------------
\begin{figure}
\centering
\includegraphics[width=\columnwidth]
{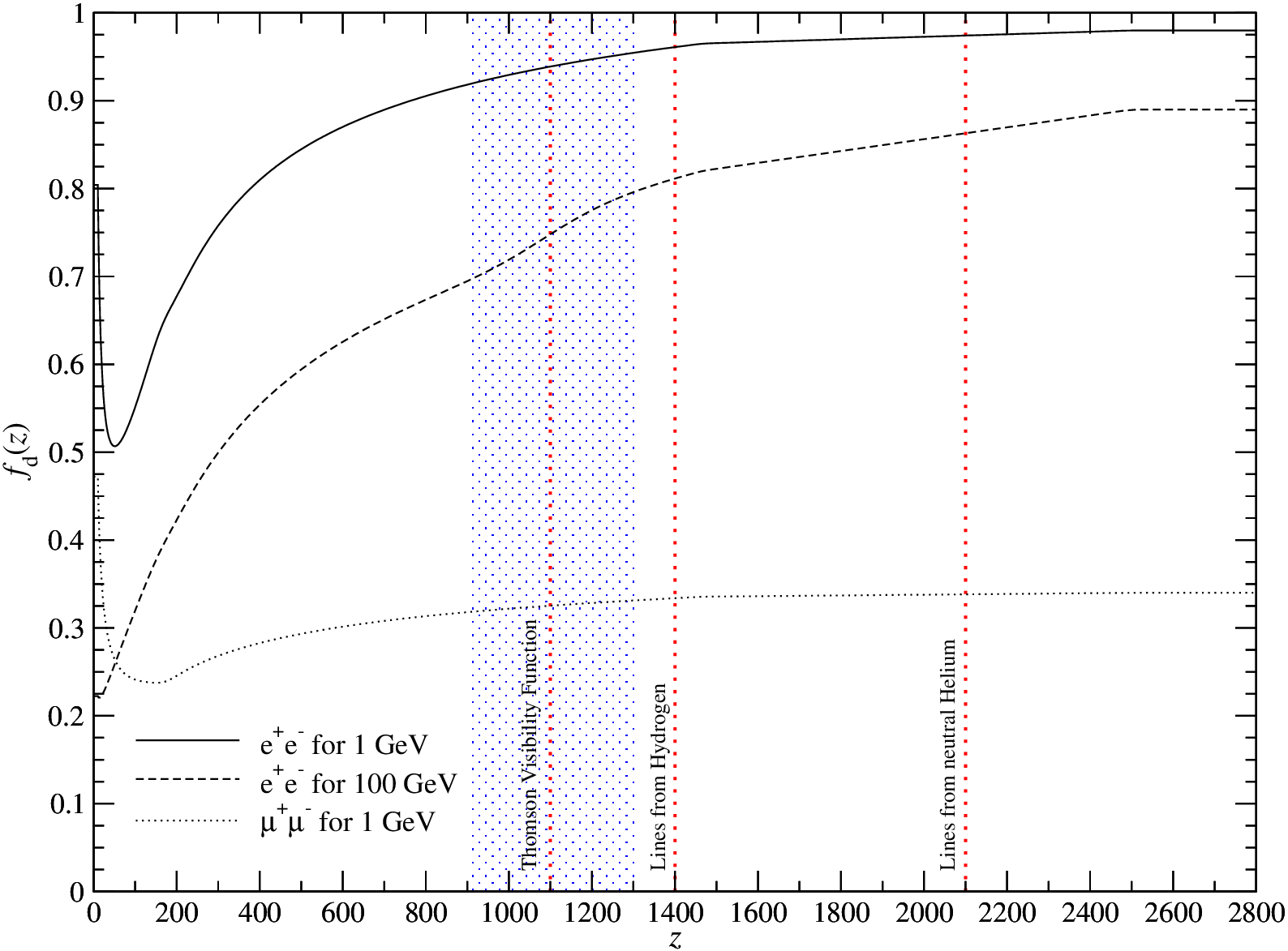}
\caption{Energy deposition efficiency, $f_{\rm d}$, for different models of annihilating DM. We also marked the region around the last scattering surface ($z\sim 1100$), and where most of the recombination lines from hydrogen ($z\sim 1400$) and neutral helium ($z\sim 2100$) are produced.}
\label{fig:fd}
\end{figure}
%---------------
%
The largest differences in the shape of $f_{\rm d}(z)$ can be found for DM models that annihilate via the channels $\chi\bar{\chi}\rightarrow e^+e^-$ with $M_{\chi}=1\,$GeV and  $M_{\chi}=100\,$GeV, and $\chi\bar{\chi}\rightarrow \mu^+\mu^-$ DM annihilation with $M_{\chi}=1\,$GeV \citep[see Fig.~4 in][]{Slatyer2009}. The functions $f_{\rm d}$ which we used to represent these cases are shown in Fig.~\ref{fig:fd}. The curves were computed applying the fitting formulae for the different redshift ranges given by \citet{Slatyer2009} and smoothly connecting them at $z\sim 170$, and linearly between $z\sim 1470$ and $z\sim 2500$.
One can see, that around the time of maximal photon production by helium ($z\sim 2200$), all of these functions are more or less constant at a level of $\sim 1\%$.
Similarly, during the time of photon release by hydrogen ($z\sim 1300-1400$) all the functions $f_{\rm d}$ are only weakly dependent on time.
The largest time-dependence is seen for the case $\chi\bar{\chi}\rightarrow e^+e^-$ with $M_{\chi}=100\,$GeV.
In all shown cases, the strongest variations appear at much lower redshifts ($z\lesssim 600$), so that from the differences in $f_{\rm d}$ one does not expect any important changes to the {\it shape} of the CMB distortions induced by the different DM annihilation models.
The dependence on the branching into heating, ionizations and excitations is much more crucial. 

The same statement applies to DM-induced modifications in the ionization history and the CMB power spectra, and was already pointed out by \citet{Slatyer2009}.
Only the overall amplitude and its conversion into DM mass and annihilation cross-section is affected.
However, since of the CMB anisotropies $f_{\rm d}(z)$ at $z\sim 1100$ matters, while for the \ion{H}{i} spectral distortions the value at $z\sim 1200-1400$ is more important, a combination of both the CMB energy spectrum and power spectra could help shedding additional light on the time-dependence of $f_{\rm d}(z)$.
Still, in the considered case the variation of $f_{\rm d}(z)$ during this time is rather small.

%---------------
\begin{figure}
\centering
\includegraphics[width=\columnwidth]
{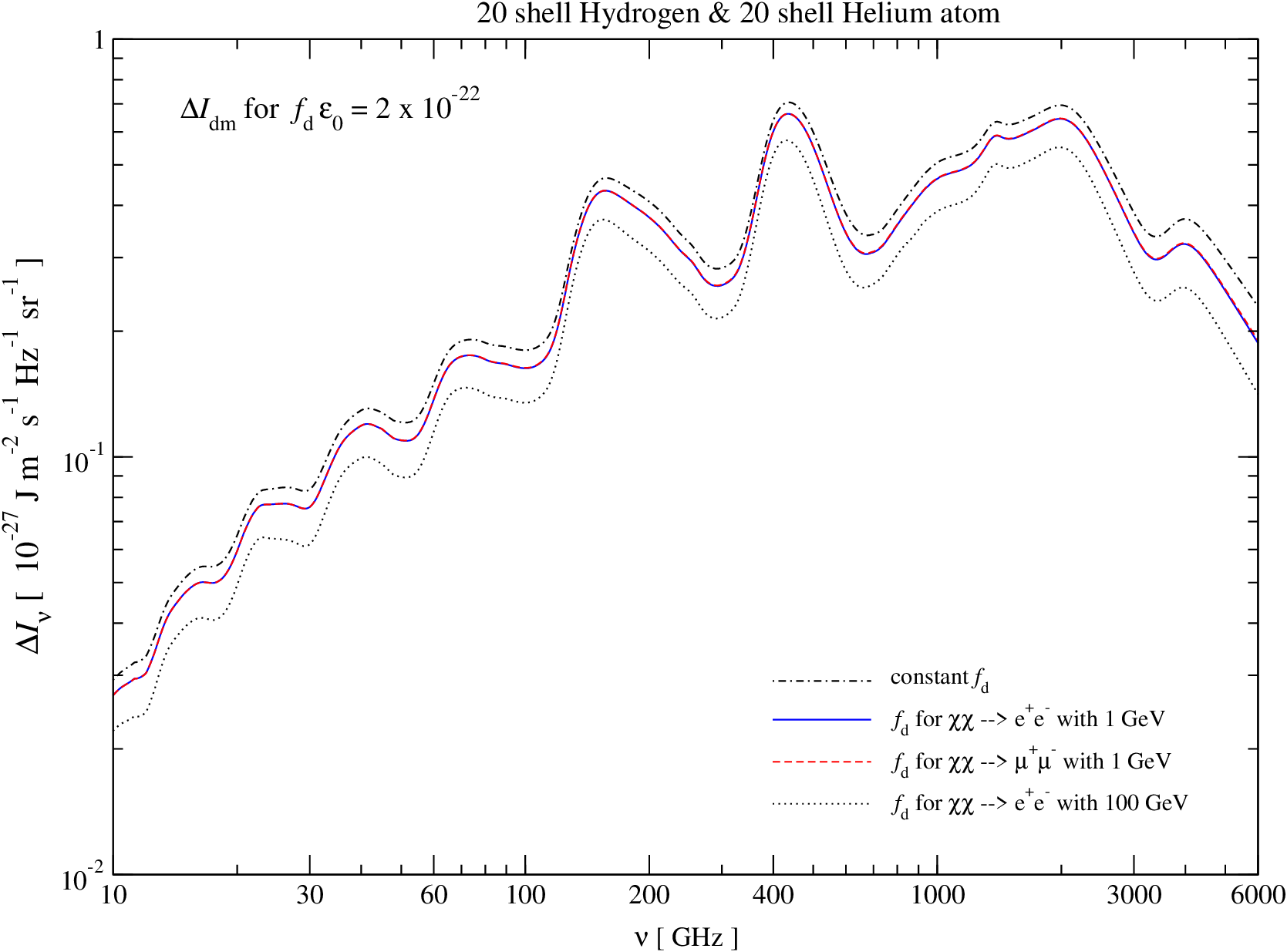}
\caption{Contribution to the total \ion{H}{i} cosmological recombination spectrum which is induced by dark matter annihilation for $f^{\rm lim}_{\rm d}\epsilon_0=\pot{2}{-22}$ and different choices of $f_{\rm d}(z)/f^{\rm lim}_{\rm d}$.
We included 20 shells for hydrogen and helium in our computations, with $\tilde{g}^{\rm a}_{\rm i}$ given by Eq.~\eqref{eq:f_i_ex_Chen}.
Like in Fig.~\ref{fig:DI_dm.HI} we give  the absolute difference, $\Delta I_{\rm dm}(\nu)$, in the {\it total}, present-day \ion{H}{i} recombination spectrum with respect to our reference model \citep{Jose2008} which does not include DM annihilations.
For the dashed-dotted line we used $f_{\rm d}(z)/f^{\rm lim}_{\rm d}=\rm const$, while for the others we used the different $f_{\rm d}(z)$ shown in Fig.~\ref{fig:fd}.
}
\label{fig:DI_dm.HI.fd}
\end{figure}
%---------------
In Fig.~\ref{fig:DI_dm.HI.fd}, we finally show the changes in the cosmological recombination spectrum for different models of annihilating DM.
To make the results comparable in amplitude, we have rescaled $f_{\rm d}(z)$ by the value at high redshift, fixing $f^{\rm lim}_{\rm d}\epsilon_0=\pot{2}{-22}$. 
One can see that the results in the cases  of $\chi\bar{\chi}\rightarrow e^+e^-$ and $\chi\bar{\chi}\rightarrow \mu^+\mu^-$ DM annihilation with $M_{\chi}=1\,$GeV are extremely similar. This is simply a consequence of the tiny differences in the redshift dependence of $f_{\rm d}(z)/f^{\rm lim}_{\rm d}$ for the two models at $z\sim 1200-1300$ (cf. Fig.~\ref{fig:fd}), from where most of the extra \ion{H}{i} distortions are coming from.
The small difference with respect to the case $f_{\rm d}(z)=\rm const$ is due to the fact that at $z\sim 1300-1400$ in both cases the value of $f_{\rm d}(z)/f^{\rm lim}_{\rm d}<1$.
This also explains the additional difference in the overall amplitude seen for the model with $\chi\bar{\chi}\rightarrow e^+e^-$ with $M_{\chi}=100\,$GeV.

If in addition we rescale the distortions so that they all coincide\footnote{We arbitrarily chose this frequency.} at $\nu\sim 200\,$GHz, then we find that  percent-level differences in all cases remain.
This shows that the model-dependence in connection with $f_{\rm d}(z)$ is very small and hence the DM-induced distortions will not be very sensitive to the total mass and magnitude of the annihilation cross-section by means of this quantity. 
This is also true for the effects on the CMB power spectra \citep[e.g.][]{Slatyer2009}, however, we expect that more detailed computations of the ionization and excitation efficiencies could reveal more pronounced dependencies on the specifics of the DM annihilation model. 

%++++++++++++++++++++++++++++++++++++++
%
%
%
%++++++++++++++++++++++++++++++++++++++
\section{Discussion}
\label{sec:disc}
%---------------
In the previous sections we have focused on illustrating the main dependencies and effects in connection with the changes that are introduced by DM annihilations during cosmological recombination.
In this section we want to go slightly beyond a purely theoretical study and include current limits on the possible DM annihilation rate into our considerations. 
We also wish to extend the discussion to more general cases of energy release and explain why the cosmological recombination spectrum could allow us to learn something about non-standard thermal histories in cases to which the ionization history and CMB power spectra are not sensitive.

\subsection{CMB power spectra versus the cosmological recombination spectrum}
\label{sec:versus}
For models which still seem to be allowed by current CMB data from {\sc Wmap} \citep{Komatsu2009}, at 95\% c.l. one has $f_{\rm d} \left<\sigma v\right>\frac{100\,{\rm GeV}}{M_\chi c^2}\lesssim \pot{3.6}{-25}\rm \,cm^3\,s^{-1}$ \citep{Galli2009, Slatyer2009}. 
In our parametrization, according to Eqs.~\eqref{eq:dEd_dt_} and \eqref{eq:deps_dt_}, this translates into $f_{\rm d} \epsilon_0\lesssim \pot{2}{-23}\,\left[\frac{\Omega_{\chi} h^2}{0.13}\right]^2$.
For this value the changes in the cosmological recombination spectrum are expected to be of the order of percent only (cf. Fig.~\ref{fig:spec.HI.A} and Fig.~\ref{fig:DI_dm.HI}). Unfortunately, this would make it rather hard to learn something in addition about DM annihilations from the cosmological recombination spectrum. So why should one try to measure the cosmological recombination spectrum when the constraints obtained with the CMB temperature and polarization power spectra are already so strong?

%---------------
\begin{figure}
\centering
\includegraphics[width=\columnwidth]
{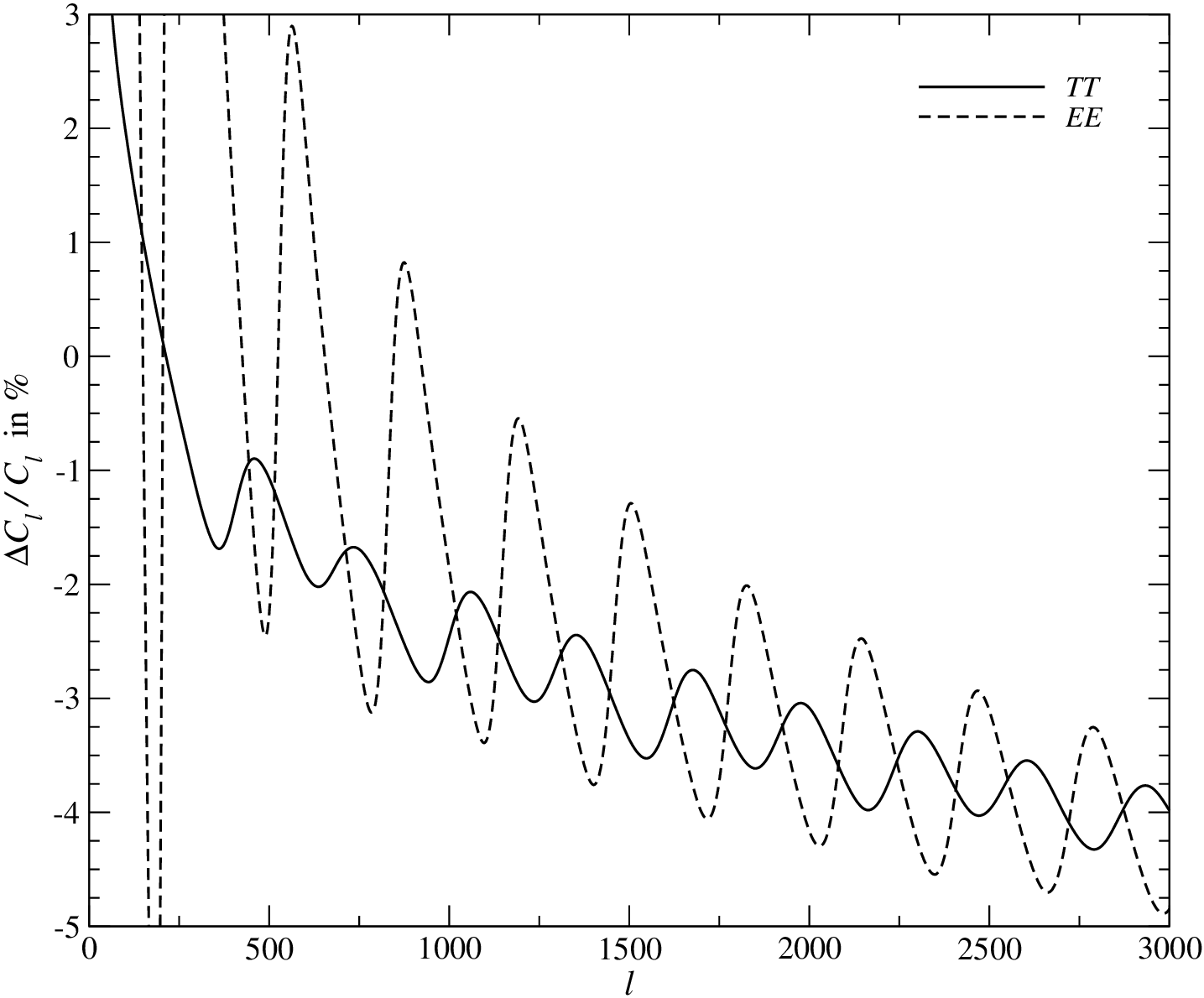}
\\[1mm]
\includegraphics[width=\columnwidth]
{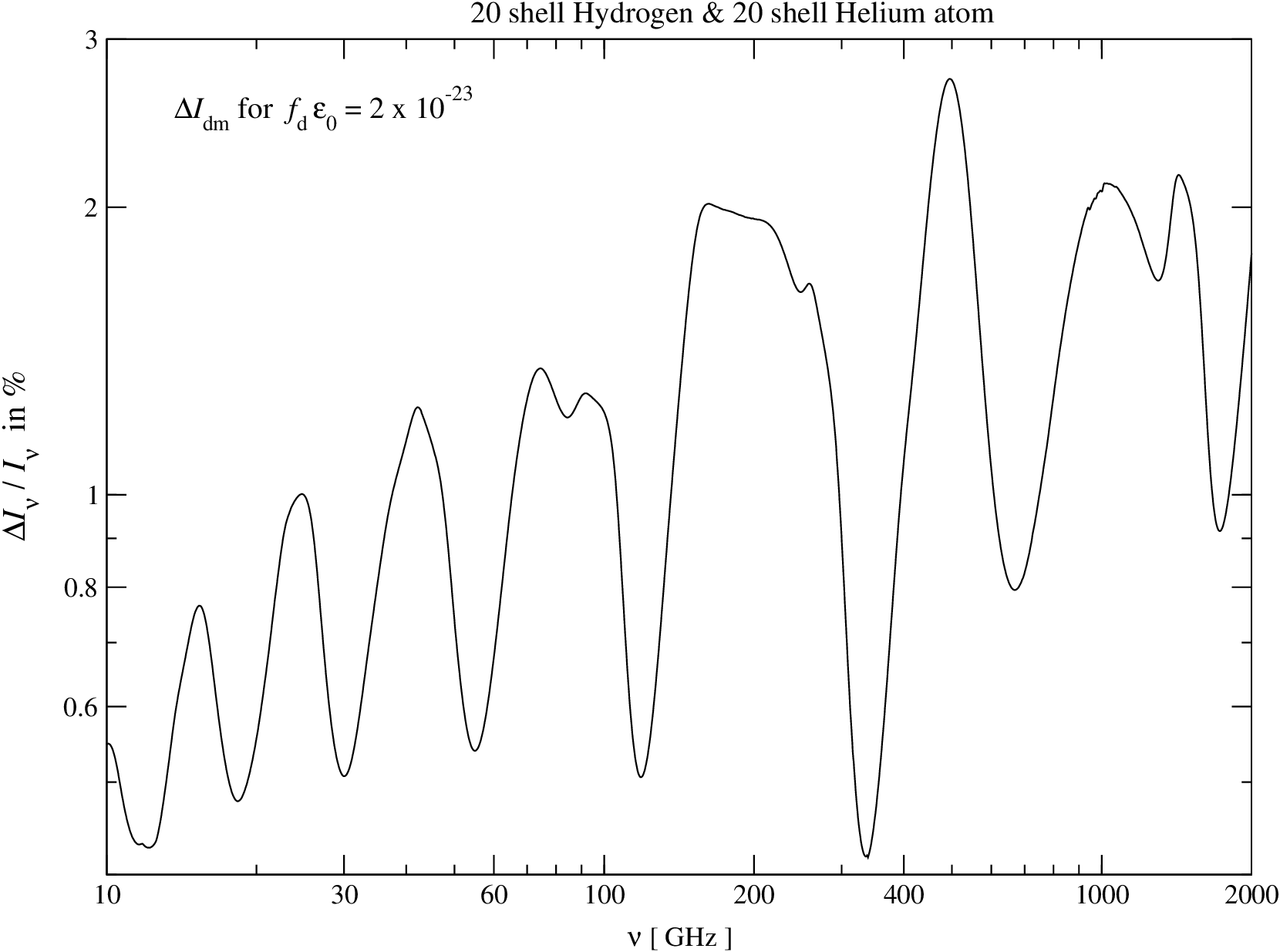}
\caption{DM-induced changes to the CMB temperature and polarization power spectra (upper panel) and the \ion{H}{i} cosmological recombination spectrum (lower panel) for $f^{\rm lim}_{\rm d}\epsilon_0=\pot{2}{-23}$.
We used {\sc Cmbeasy} to compute the changes in the CMB power spectra \citep{Doran2005}.
}
\label{fig:DCl}
\end{figure}
%---------------
%
First of all, one should emphasize that for the allowed values of $f_{\rm d} \left<\sigma v\right>$ also the changes in the CMB temperature and polarization power spectra are fairly small: For $f_{\rm d} \left<\sigma v\right>=\pot{2}{-23}$ they reach $\sim 4\%-5\%$ at high multipoles (see Fig.~\ref{fig:DCl}), where the main trend is completely featureless, and the variable part has an amplitude of only $\sim 1\%$.
In this context it is particularly important that due to {\it cosmic variance} percent-level correction to the CMB power spectra only become statistically significant at high multipoles.

Secondly, it is clear that the DM-induced modifications in the CMB power spectra are strongly {\it degenerate} with the values of the scalar spectral index $n_{\rm s}$, the baryon density $\Omega_{\rm b}$, and $\sigma_{8}$ \citep[e.g. see][]{Padmanabhan2005, Slatyer2009, Galli2009}. 
Here also previously neglected physical processes \citep[see][for overview]{Fendt2009,Sunyaev2009} are important, since for {\sc Planck} they can lead to significant biases to the values of $n_{\rm s}$ and $\Omega_{\rm b}$ \citep{Jose2009}, or otherwise another {\it confusion} to the possible signatures of DM annihilations.

Furthermore, at high multipoles the signals related to SZ-clusters \citep[e.g. see][]{Molnar2000, Refregier2000, Holder2001, Komatsu2002} and experimental systematics \citep[e.g. due to calibrations of the beam,][]{Colombo2009} become very important. 
Disentangling all these components is a mayor problem, and this is where the cosmological recombination spectrum could help in addition:
just looking at the amplitude of the DM-induced changes the cosmological recombination spectrum, it is clear that in this respect the recombination spectrum in principle is similarly sensitive as the CMB power spectra (see Fig.~\ref{fig:DCl}). 
Also, the cosmological recombination spectrum {\it does not} depend on the value of the scalar spectral index $n_{\rm s}$, so that such confusion is already excluded.
In addition, the cosmological recombination spectrum is not limited by {\it cosmic variance}: no statistical comparison of the measured energy spectrum with some ensemble of Universes is involved. One would investigate the recombination spectrum for our particular realization of the Universe, only encountering small fluctuations of the cosmological parameters in different directions of the sky. It even is possible to take these corrections into account in the computations of the cosmological recombinations spectrum, but for standard cosmological model the effects are expected to introduce changes that are smaller than $\Delta I_\nu/I_\nu \sim 10^{-4}-10^{-3}$.

To observe the cosmological recombination spectrum no absolute measurement is necessary \citep{Chluba2008T0, Sunyaev2009}. Due to its very peculiar frequency-dependence, it should therefore in principle be possible to separate the cosmological recombination spectrum from foregrounds and instrumental signals.
One can use particularly clean patches on the sky, observing with wide-angle horns. Also the cosmological recombination signal should be polarized at a very small level only, providing another possibility to discriminate it from other signals.
By measuring the positions, and width of several spectral features in the cosmological recombination spectrum with high precision one should therefore in principle be able to extract valuable information on possible sources of extra ionizations and excitation during recombination.

\subsection{Why the cosmological recombination spectrum could teach us something in addition}
\label{sec:Why}
%---------------
Until now we have only considered the possibility of continuous, very extended energy release caused by DM annihilations.
However, as an example, for energy injection due to {\it long-lived decaying particles} additional aspects become important:
First of all, in these cases most of the energy will be released over a characteristic time $\Delta z/z\sim 30\%-40\%$. This implies that for decaying particles the possible changes to the ionization history or cosmological recombination spectrum could be significantly more sharp. 
Secondly, if the life-time of the particle is shorter than $\sim 380\,000$ years, then most of the energy will be released before the maximum of the Thomson visibility function. In that case the effect on the CMB temperature and polarization power spectra could be much smaller or even negligible, while the changes in the cosmological recombination spectrum would still be very strong.

For example, if energy were released at $\sim 18\,000$ years after the big bang (corresponding to the time of $\ion{He}{iii}\rightarrow \ion{He}{ii}$ recombination) then the CMB power spectra would practically remain unchanged, while at the same time due to extra ionizations and excitations the number of photons produced by \ion{He}{ii} could increase significantly\footnote{Such behavior was seen in connection with the possibility of pre-recombinational energy release, where the contributions from \ion{He}{ii} to the recombination radiation became as large as those from hydrogen \citep{Chluba2008c}.
However, we note that in the case considered here, there is no significant $y$-type distortion produced by the energy injection, but as mentioned in the introduction, one can still have a very important effect on the cosmological recombination spectrum, since we allow for {\it direct} ionizations and excitations of neutral atoms.}.
Similarly, if the energy is released between the two epoch of helium recombination or during the recombination of neutral helium, the CMB power spectra will mostly be unaltered, while a lot of extra emission by \ion{He}{ii} could be induced.
The cosmological recombination spectrum therefore may allow us to check for extra sources of ionizations or excitations at times where the CMB temperature  and polarization power spectra are not sensitive.

It is also important to mention, that any other modification of the recombination process, e.g. caused by the {\it variation of fundamental constants} like the fine-structure constant $\alpha$, or Newtons gravitational constant $G$ \citep[see][for references]{Scoccola2009, Galli2009b}, will also lead to changes in the shape and positions of features in the cosmological recombination spectrum. Observing such changes may therefore provide another very clean and direct way to test non-standard physics at early stages of the Universe.

\subsection{Towards detailed templates for the cosmological recombination spectrum}
There is no principle difficulty in computing the cosmological recombination spectrum with $\sim 0.1\%$ accuracy, also including the effects of possible energy or particle release in the recombination epoch, even for more general cases.
However, in this case, one would have to take other previously neglected physical processes \citep[see][for overview]{Fendt2009, Sunyaev2009} into account in addition. Here in particular those connected with the Lyman $\alpha$ radiative transfer \citep{Chluba2008b, Chluba2009b, Hirata2009} and two-photon transitions \citep{Chluba2008a, Hirata2008, Chluba2009} will be very important, because they should affect the recombination spectrum at the level of $\sim 10\%$ \citep[see comments in][]{Chluba2009b}. 
Also the effect of electron scattering will be important \citep{Jose2008, Chluba2008d}, in particular for the contributions from helium.

Furthermore, the computations of the heating, ionization and excitation efficiencies will also have to be refined in order to study detailed model-dependencies (for additional comments see Sect.~\ref{sec:ion_ex}), and as we have seen in Sect.~\ref{sec:DM_ann_Spec_const_feps_g}, the difference in the recombination spectrum can be large.
Also one should include more shells for hydrogen and helium into the computations, since the total amplitude of the additional distortions is still expected to increase, especially at low frequencies.

In addition, the emission due to \ion{He}{ii} should be taken into account. Given that there is a very extended period between the recombination of \ion{He}{ii} at $z\sim 6000$ and \ion{He}{i} at $z\sim 2200$, the total amount of \ion{He}{ii} emission induced by DM annihilations is expected to be very important, possibly even exceeding that from neutral helium.
Such calculations are beyond the scope of this paper, but we plan to investigate these aspects in more detail in the future.

However, we note that currently it will probably be more important to investigate more general observational prospects in connection with a measurement of the cosmological recombination spectrum, including possible foregrounds and systematics. Also, it will be very important to understand, which frequency bands will be most useful and sensitive to changes in the cosmological parameters or energy injection.
It is obvious that measuring the cosmological recombination spectrum will be very challenging, but on the other hand there could be a lot to learn from this.
In particular, a combination of the CMB temperature and polarization anisotropies with the CMB energy spectrum could open a way to further tighten CMB-based constraints on standard and non-standards aspect of our cosmological model.

%++++++++++++++++++++++++++++++++++++++
%
%
%
%++++++++++++++++++++++++++++++++++++++
\section{Conclusions}
\label{sec:conc}
%---------------
Assuming that DM is annihilating throughout the history of our Universe, we have demonstrated that the cosmological recombination spectrum in principle is sensitive to the branching of the deposited energy into heating, ionizations and excitations.
If energy only goes into heating of the medium (without leading to some significant primordial $\mu$ or $y$-type CMB distortion), the recombination spectrum in practically not affected, while extra ionizations and excitations lead to modifications in both the contributions from hydrogen (cf. Fig.~\ref{fig:spec.HI.A}) and helium (cf. Fig.~\ref{fig:spec.HeI.A}).

We have shown, that the overall amplitude of the DM-induced spectral distortions depends on the total amount of ionizations and excitations (Fig.~\ref{fig:DI_dm.HI}) at $z\sim 1200-1300$ for hydrogen, and $z\sim 2000-2400$ for helium.
Furthermore, the relative importance of DM-induced excitations and ionizations, and the time-dependence of their efficiencies determine the exact {\it shape} and {\it position} of the additional CMB spectral distortions from the recombination epoch (see Fig.~\ref{fig:DI_dm.HI.ex}).
Since these efficiencies depend on the considered model for the annihilating DM, or more generally, the process that produced the additional ionizations and excitations (e.g. decaying particles), by measuring the cosmological recombination spectrum in several spectral bands in principle one should be able to place additional constraints on possible energy release during cosmological recombination.
Given the rather strong dependence of both the changes to the cosmological recombination spectrum and the CMB temperature and polarization power spectra on these efficiencies, for precise predictions in connection with possible extra release of energy during recombination it will be important to refine the computations of these efficiencies in the cosmological context (see Sect.~\ref{sec:ion_ex} for more comments). 

Although for currently allowed values of the effective DM annihilation rate the changes to the cosmological recombination spectrum are of the order of percent, we have argued that there are several reasons to believe that one could learn something in addition by studying the signals from the recombination era (see Sect.~\ref{sec:disc}). In particular, the cosmological recombination spectrum is expected to be sensitive to cases of energy or particle release (e.g. before the maximum of the Thomson visibility function), by which the CMB temperature and polarization power spectra are not affected (see Sect.~\ref{sec:Why}).
Beyond a check that recombination has occurred as we think it has, observing the cosmological recombination radiation would therefore allows us to directly check our understanding of {\it standard} and {\it non-standard} physical processes happening at about $260\,000$, $130\,000$ and $18\,000$ years after the big bang. 
%
%Combining such measurements with those for the CMB temperature and polarization power spectra and other cosmological data sets, should allow us to %break degeneracies, for example in connection with the scalar spectral index $n_{\rm s}$ (see comments in Sect.~\ref{sec:versus}).

\section*{Acknowledgements}
JC is grateful to R.A. Sunyaev for useful discussions and suggestions.
He would also like to thank D. Giannios, G. H{\"u}tsi, C.~Pfrommer, J.~A.~Rubi{\~n}o-Mart{\'{\i}}n and C.~Thompson for their comments.
Futhermore, he would like to acknowledge the use of computational resources at MPA.

\begin{appendix}

%++++++++++++++++++++++++++++++++++++++
%
%
%
%++++++++++++++++++++++++++++++++++++++
%\section{Simple estimates regarding the problem}
%\label{sec:estimates}
%%---------------

\section{Estimates for the primordial CMB spectral distortions for the considered models of annihilating DM}
\label{sec:est_y}
%---------------
A significant fraction of the total energy released due to DM annihilation is going into heating of the medium, in particular at redshifts well before the recombination epoch, where $f_{\rm h}\approx f_{\rm d}$ and $f_{\rm ion}\approx f_{\rm ex}\approx 0$.
It is clear that this will lead to some (tiny) primordial distortion of the CMB, since at redshifts $z\lesssim \pot{2}{6}$ the thermalization process stops being 100\% efficient \citep[e.g.][]{Illarionov1975b, Burigana1991, Hu1993b}. 

One can estimate the expected {\it primordial} distortion of the CMB by computing the total change in the energy density of the CMB, 
$\Delta \rho_\gamma/\rho_\gamma=\int \frac{\Delta \dot{\rho}_\gamma}{\rho_\gamma} \id t$, that is caused by the deposition of energy in the DM annihilation process.
%%-----------
%\beal
%\label{eq:drho_rho}
%\Delta \rho_\gamma/\rho_\gamma=\int \frac{\Delta \dot{\rho}_\gamma}{\rho_\gamma} \id t.
%\end{align}
%%-----------
For a $\mu$ and $y$-type spectral distortion one then expects $\mu\approx 1.4\,\Delta \rho_\gamma/\rho_\gamma$ and $y\approx \frac{1}{4}\Delta \rho_\gamma/\rho_\gamma $ \citep[e.g.][]{Illarionov1975b}.
For $s$-wave annihilation ($\left<\sigma v\right>=\rm const$) one finds \citep[cf.][]{McDonald2001}
%-----------
\beal
\label{eq:mu_est}
\mu\approx \pot{3}{-10}\,
f_{\rm d, lim}
\left[\frac{M_\chi c^2}{100\,{\rm GeV}}\right]^{-1}
\left[\frac{\Omega_{\chi} h^2}{0.13}\right]^2 
\left[\frac{\left<\sigma v\right>}{\pot{3}{-26}\,{\rm cm^3/s}}\right],
\end{align}
%-----------
where $f_{\rm d, lim}$ is the limiting energy deposition efficiency at large $z$.
This is only a rough estimate, but since the final value for $\mu$ is so far below the current upper limit $|\mu|<\pot{9}{-5}$ obtained with {\sc Cobe/Firas} \citep{Mather1994, Fixsen2002}, for our purpose it is sufficient.
The effective $y$-parameter \citep{Zeldovich1969} is of the same order. 

Although for $y\sim 10^{-7}-10^{-6}$ a significant amount of photons is generated in the {\it pre-recombinational epochs} of hydrogen and helium \citep{Chluba2008c}, inducing interesting narrow spectral features in the cosmological recombination radiation, at the level of $\mu$ or $y\sim 10^{-10}$ this process can be completely neglected.
In our computations we can therefore assume that the ambient CMB radiation field is given by a blackbody with present-day temperature $T_0=2.725\,$K \citep{Fixsen2002}.

\end{appendix}

\bibliographystyle{mn2e} 
\bibliography{Lit}

\end{document}